\documentclass[%
 reprint,
 amsmath,amssymb,
 aps,
prb,
]{revtex4-1}

\usepackage{graphicx}% Include figure files
\usepackage{dcolumn}% Align table columns on decimal point
\usepackage{bm}% bold math
\usepackage{color}
\usepackage[super]{nth}
\usepackage{float}
\usepackage{subfigure}
\usepackage{multirow}
\usepackage{dcolumn}
\usepackage{gensymb}
\usepackage{bm}
\usepackage{epstopdf}
\usepackage{siunitx}
\usepackage{subfigure}
\usepackage{amsmath}
\usepackage{color}
\usepackage{units}
\usepackage{natbib}
\usepackage{xpatch}
\usepackage{scrextend}
\usepackage{graphicx}% Include figure files
\usepackage{dcolumn}% Align table columns on decimal point
\usepackage{bm}% bold math
\usepackage{hyperref}
\usepackage{enumitem}
\usepackage{array}
\usepackage{etoolbox} % for \appto
\usepackage{lipsum} % for mock text
\usepackage[capitalize]{cleveref}

\makeatletter
\newcolumntype{P}[1]{>{\centering\arraybackslash}p{#1}}
\appto{\appendix}{%
  \@ifstar{\def\theequation@prefix{A.}}%
          {}%
}
\makeatother

\hypersetup{colorlinks,linkcolor={black},citecolor={blue},urlcolor={blue}} 

\apptocmd{\thebibliography}{\raggedright}{}{}

\begin{document}

%\preprint{APS/123-QED}

\title{Defect energetics of cubic hafnia from quantum Monte Carlo simulations}% Force line breaks with \\

\author{Raghuveer Chimata}
\affiliation{Argonne Leadership Computing Facility, Argonne National Laboratory, Lemont, Illinois 60439, USA}

\author{Hyeondeok Shin}
%\affiliation{Argonne Leadership Computing Facility, Argonne National Laboratory, Lemont, Illinois 60439, USA}

\author{Anouar Benali}
%\affiliation{Argonne Leadership Computing Facility, Argonne National Laboratory, Lemont, Illinois 60439, USA}
\affiliation{Computational Sciences Division, Argonne National Laboratory, Lemont, Illinois 60439, USA}

\author{Olle Heinonen}
\affiliation{Material Science Division, Argonne National Laboratory, Argonne, Illinois 60439, USA}
\affiliation{Northwestern-Argonne Institute for Science and Engineering, 2205 Tech Drive, Suite 1160, Evanston, Illinois 60208, USA}

\email[Corresponding author:]{heinonen@anl.gov}

\date{\today}% It is always \today, today,

\begin{abstract}
Cubic hafnia (HfO$_2$) is of great interest for a number of applications in electronics because of its high dielectric constant. However, common defects in such applications degrade the properties of hafina. We have investigated the electronic properties of oxygen vacancies and nitrogen substitution in cubic HfO$_2$ using first principles calculations based on density functional theory (DFT) and many-body diffusion Monte Carlo (DMC) methods. We investigate five different charge defect states of oxygen vacancies, as well as substitutional N defects which can lead to local magnetic moments. 
Both DMC and DFT calculations shows that an oxygen vacancy induces strong lattice relaxations around the defect. Finally, we compare defect formation energies, charge and spin densities obtained from DMC with results obtained using DFT. While the obtained formation energies from DMC are 0.6~eV -- 1.5~eV larger than those from GGA+U, the agreement for the most important defects, neutral and positively charged oxygen vacancies, and nitrogen substitutional defect, under oxygen-poor conditions are in reasonably good agreement. Our work confirms that nitrogen can act to passivate cubic hafnia for applications in electronics. 
\end{abstract}

\pacs{Valid PACS appear here}% PACS, the Physics and Astronomy
\maketitle

%\tableofcontents

\section{\label{sec:level1} Introduction}
The continuous progress in miniaturization of electronic devices such as silicon-based complementary metal-oxide-semiconductor (CMOS) field effect transistor (FET) has let to the replacement of silica (SiO$_2$) as the gate dielectric material by materials with dielectric constants, $k$, higher than that of silica~\cite{doi:10.1063/1.126214,doi:10.1063/1.1611644,ROBERTSON20151}. The use of high-$k$ dielectrics as gate materials dramatically reduces the gate leakage current due to electron-tunneling. High-$k$ materials can be used as ultra-thin layers ($\sim$1~nm), but have similar gate dielectric properties as regular silica layers. Among the possible choices of high dielectric constant materials for CMOS, hafnia has become an attractive material due to its favorable properties such as the wide band gap of 5.25 -- 5.95~eV \cite{PhysRevB.81.085119} and high $k$ value of 22 \cite{doi:10.1063/1.4878401}. Besides, hafnia has excellent chemical compatibility with silicon, and a higher heat of formation than silica. Moreover, hafnia is thermally and chemically stable. This is especially important for the silica contact because gate stacks undergo a rapid thermal annealing processes. Due to these extraordinary features, amorphous hafnia was introduced as a dielectric material in CMOS devices by Intel over a decade ago\cite{Huang10}. Current technology can achieve a $\sim$1~nm equivalent oxide thicknesses with high-$k$ in Si-CMOS devices, and the next step is to find higher-$k$ dielectrics ($k > 22$) with a band offset compatible with silica. 

Hafnia can exist in three polymorphic phases at atmospheric pressure: monoclinic (T $<1700$~K), tetragonal ($1700$~K $<$ T $<2600$~K ) and cubic (T $>$ 2600~K). First principles studies~\cite{PhysRevB.65.233106} have shown that the tetragonal ($k \sim$ 70) and cubic ($k \sim$ 29 ) phases have a much larger dielectric response than the monoclinic phase ($k\sim$ 16). Incorporation of lanthanides has been shown to stabilize the high-temperature phases of hafnia  \cite{doi:10.1063/1.1880436,doi:10.1063/1.2798498,doi:10.1063/1.2216102}. Recently, S. Migita et al., have demonstrated that ultra-thin cubic hafnia films exhibit a very high-{$k$} value of about 50 and have band gap similar to that of  monoclinic hafnia.  Ultrathin films of cubic hafnia have been demonstrated using an ultra-fast ramp with a shorter hold time in the annealing process from as-deposited amorphous hafnia \cite{4588599}. Other, more recent works, have also demonstrated low-temperature synthesis routes to highly crystalline cubic hafnia\cite{Kumar2017}.

However, depending on the particular deposition process, hafnia has different material properties and often has a reduced dielectric constant; in addition, leakage currents in thin film hafnia can also pose a problem in gate dielectric applications. A lower than ideal performance in electronic gates can be explained by the formation of defect-related fixed charges \cite{1196025,doi:10.1063/1.1626019}. The monoclinic phase has been well studied both experimentally via scattering techniques \cite{PhysRevB.90.134426,doi:10.1063/1.4901961}, and theoretically using     DFT\cite{PhysRevB.75.104112,XIONG2005408,PhysRevB.65.174117,PhysRevB.81.085119}. Moreover, first-principles calculations carried out on the low-temperature monoclinic phase of hafnia with oxygen vacancies and oxygen interstitials suggest that the oxygen vacancies represent the main electron traps \cite{XIONG2005408}. A quantitative analysis of electronic states associated with strongly lattice-coupled localized oxygen defects in a cubic hafnia has been described by the negative-{U$^{\prime}$} Anderson model\cite{doi:10.1063/1.2009826}, while the incorporation of nitrogen into silica~\cite{doi:10.1063/1.118389} and hafnia \cite{Maurya2018} has been shown to reduce the gate leakage currents.

In addition to applications as a gate dielectric, hafnia-based resistive random access memory (RRAM) devices exhibit excellent switching characteristics and reliable data retention which makes them useful as non-volatile devices. However, the switching performances of these devices can be greatly affected by charged oxygen defects\cite{doi:10.1063/1.3694045,doi:10.1063/1.3543837}. Recently, oxygen-modulated quantum conductance for ultrathin hafnia-based memristive switching devices was studied using DFT-based quantum transport simulations~\cite{doi:10.1063/1.3694045,PhysRevB.94.165160}. However, accurate studies of oxygen defects in hafnia are still missing.

Defects in hafnia can give rise to other phenomena. Unexpected ferromagnetism has been observed in HfO$_{2}$ thin films \cite{Venkatesan2004,PhysRevB.72.024450}, and a first-principles study has shown that Hf vacancies could be the possible origin of the ferromagnetism \cite{PhysRevLett.94.217205}. In a very similar system, ZrO$_{2}$, first-principles calculations have shown that by doping with nitrogen, the system becomes ferromagnetic. The reported total magnetic moment is 1.0~$\mu_{B}$ per N defect. In contrast, a study by Hildebrandt et al.\cite{PhysRevB.90.134426}, taking into account a broad range of oxygen vacancy concentrations and magnetic dopants, has shown that undoped, oxygen-deficient, or magnetically doped hafnia does not possess intrinsic ferromagnetism.

Several theoretical studies on the defect formation energies and energy levels in HfO$_{\rm 2}$ have been carried out using density functional theory (DFT) with different functionals and basis functions\cite{PhysRevB.75.104112,XIONG2005408,PhysRevB.65.174117,doi:10.1021/acs.jpcc.6b06913}. The quality and consistency of the calculated energetics, such as defect formation energy, vary on quite a large scale, 0.04~eV -- 0.27~eV, between different DFT exchange-correlation functionals. This large variation may be related to the fact that hafnia is a correlated material, and the $5d$ localized electrons should be treated with methods that can take these correlations into account. In order to accurately address the problem of the correlated $5d$ electrons, we use Quantum Monte Carlo (QMC) methods, in particular, diffusion Monte Carlo (DMC)~\cite{RevModPhys.73.33,doi:10.1063/1.443766}, to compute the ground state electronic structure properties of this material. DMC is a stochastic sampling method to solve the many-body Schr\"{o}dinger equation\cite{RevModPhys.73.33}. It is a powerful computational technique that has provided highly accurate many-body {\it ab-initio} simulations of solids with no empirical parameters~\cite{0953-8984-30-19-195901}.  Addressing strongly-correlated systems using DMC has demonstrated accuracy and required only few controlled approximations~\cite{doi:10.1063/1.4919242,PhysRevMaterials.1.073603}.
Most previous studies have been performed on the monoclinic phase, while the tetragonal and cubic phases have been much less studied. In addition, current experimental data do not provide a detailed fundamental understanding of oxygen-deficient and doped cubic hafnia, which is of much interest for future ultra-thin CMOS applications. Our study is aimed at addressing this gap in fundamental knowledge. 

The article is organized as follows: In Sec.~\ref{method} we briefly describe the theoretical framework and computational approach employed in this work. In Sec.~\ref{results} we discuss the electronic properties of hafnia with different oxygen vacancy charge states and nitrogen dopants, and we compare charge and spin densities obtained within the Generalized Gradient Approximation (GGA)\cite{PhysRevLett.77.3865}  of DFT with a Hubbard-U added to the Hf $5d$ orbitals to account for on-site Coulomb repulsion~\cite{PhysRevB.44.943} with charge densities computed by DMC. Sec.~\ref{conclusion} summarizes the main findings of this work.

\section{Method}\label{method}
Electronic structure calculations were performed using the fixed-node diffusion Monte Carlo method as implemented in the QMCPACK code~\cite{1742-6596-402-1-012008}. We used a standard single-determinant Slater-Jastrow\cite{Jastrow1955} trial wave function with one-, two-, and three-body Jastrow functions describing the  ion-electron, electron-electron and ion-electron-electron correlations, respectively. The two-body\cite{Fahy90} and three-body Jastrow functions\cite{Drummond04} are spin dependent and coefficients for all one-, two- and three-body Jastrows were optimized using VMC. The form of the Jastrows used in this paper are described in Ref.~[\onlinecite{1742-6596-402-1-012008}].   

DFT single-particle Kohn-Sham (KS) orbitals were used to generate single Slater determinant trial wave functions for the QMC calculations. The KS orbitals were generated using plane wave (PW) basis sets with the QUANTUM ESPRESSO code\cite{0953-8984-21-39-395502}.
In this study, a scalar-relativistic pseudopotential for Hf was generated using the OPIUM package within a plane-wave basis set~\cite{OPIUMPackageNoTitle}. The core-valence interactions were treated through norm-conserving pseudopotentials and semicore states included into valence electrons. We used the following valence electronic configurations for the hafnium, oxygen, and nitrogen atoms, respectively: [Pd + 4f$^{14}$]5s$^2$5p$^6$5d$^2$6s$^2$ (Hf), [He]2s$^2$2p$^4$ (O) and  [He]2s$^2$2p$^3$ (N). The exchange-correlation potential was treated using the GGA with the Perdew, Burke, and Ernzerhof (PBE) functional~\cite{PhysRevLett.77.3865} and an on-site Hubbard U correction\cite{PhysRevB.57.1505} applied to the Hf $5d$ electrons. We used the cubic structure for hafnia with a supercell size of 96 atoms. The self-consistent DFT calculations were computed with a $4\times4\times4$ k-point mesh and a 450~Ry plane-wave kinetic energy cut-off. The optimized U-parameter used in this study was U = 2.2~eV for the Hf $5d$ orbitals. This value of U was obtained from DMC calculations (see appendix~\ref{sec:appen}). In a previous study\cite{PhysRevMaterials.2.075001},  the DMC calculated lattice parameter was found to be 5.04(1){\AA} for a cubic-HfO$_2$, while the experimentally measured values at high temperatures were extrapolated to 5.08~{\AA} at 0~K.  As a reasonable estimate for the room-temperature lattice constant, we used a compromise value of 5.07~{\AA}.

Electronic structure studies of point defects in cubic hafnia are very relevant to understanding the energetics of defect formation and the stability of the defects. We created vacancy or substitutional defects in our supercell by removing an oxygen atom, leaving an oxygen vacancy ($V_O$) behind, or by removing an oxygen atom and replacing it by a substitutional defect, such as N. Also, defect states can have different charges, which we considered. For example, there are five possible charge states for an oxygen vacancy, namely V$_{\rm O}^{\rm -2}$, V$_{\rm O}^{\rm -1}$, V$_{\rm O}$, V$_{\rm O}^{\rm +1}$ and V$_{\rm O}^{\rm +2}$.  Once the defect was created, we fully relaxed the positions of all atoms while keeping the supercell lattice vectors fixed until the force acting on each atom was less than 0.0004 eV/{\AA}.\\

In general, the formation energy $E_f$ for a point defect with charge $q$ is,  
 \begin{equation}\label{eq2}
E_{f}({\mu_{i},E_{F}})=E^{q}_{\rm def}-E_{\rm ideal}+\sum_{i}\mu_{i}+q(E_{\rm VBM}+E_{\rm F})
\end{equation}
where E$^{q}_{\rm def}$ is the total energy of the defective supercell with corresponding charge state $q$, $E_{\rm ideal}$ is the total energy of the ideal supercell, $\mu_{i}$ is the elemental chemical potential with a positive sign for vacancies and negative sign for substitutional defects; E$_{\rm VBM}$ is the valence band maximum of the ideal bulk system, and $E_{\rm F}$ is a Fermi level. In this study, we reference $E_{\rm F}$ with respect to the valence-band maximum E$_{\rm VMB}$. In order to assess the thermodynamic stability of different oxygen defect states in hafnia we need  the chemical potential of oxygen $\mu_{\rm O}$. The chemical potential of oxygen can be obtained under two conditions, oxygen-rich and oxygen-poor. Under oxygen-rich conditions, the oxygen chemical potential $\mu_{\rm O}$ is computed as $\mu_{\rm O} \approx \frac{1}{2} E({\rm O}_2)$ , where $E({\rm O}_{\rm 2}$) is the total energy of an oxygen dimer. The computed chemical potential of oxygen, $\mu_{\rm O}$ is 871.36~eV (GGA) and 869.45(3)~eV (DMC). For oxygen-poor conditions, the oxygen chemical potential is instead obtained as  $\mu_{\rm O} =  (\mu_{\rm Hf} -\mu_{\rm HfO_{2}})/2$. The main sources of errors in our QMC calculations are the finite time-step and finite system size. In order to mitigate errors, we used a small time-step of 0.005 Ha$^{\rm -1}$ and a supercell size of 96 atoms. One-body finite-size effects~\cite{PhysRevE.64.016702} in the periodic supercell DMC calculations were reduced by using twist-averaged boundary conditions; 
here energies were converged using a $2\times2\times2$ twist grid, reduced to four high-symmetry twists.
While two-body effects can only be fully corrected by extrapolating the supercell size to infinite\cite{Leslie_1985,Gillan2005}, this procedure can only work for pure systems (without defects). When studying impurities in a bulk, the concentration of impurities must remain constant and finite size extrapolation become tedious. We have analyzed two-body finite-size errors in the ionization potential (IP) using a 96-atom cell and  a 192-atom cell. The calculated IPs are 10.3(2) and 10.2(4) eV, respectively. While it is impossible to extrapolate accurately to the infinite-size limit using only two data points,  both IPs are within each other's error bars suggesting a small dependence of the energy on the two-body corrections. These two cell sizes used  modified periodic Coulomb interactions~\cite{PhysRevB.78.125106} and Chiesa-Ceperley-Martin-Holzmann kinetic energy~\cite{PhysRevLett.97.076404} corrections.

\section{Results}\label{results}
\subsection{Relaxation of ions}
\begin{figure}[!ht]
\includegraphics[width=1.0\columnwidth]{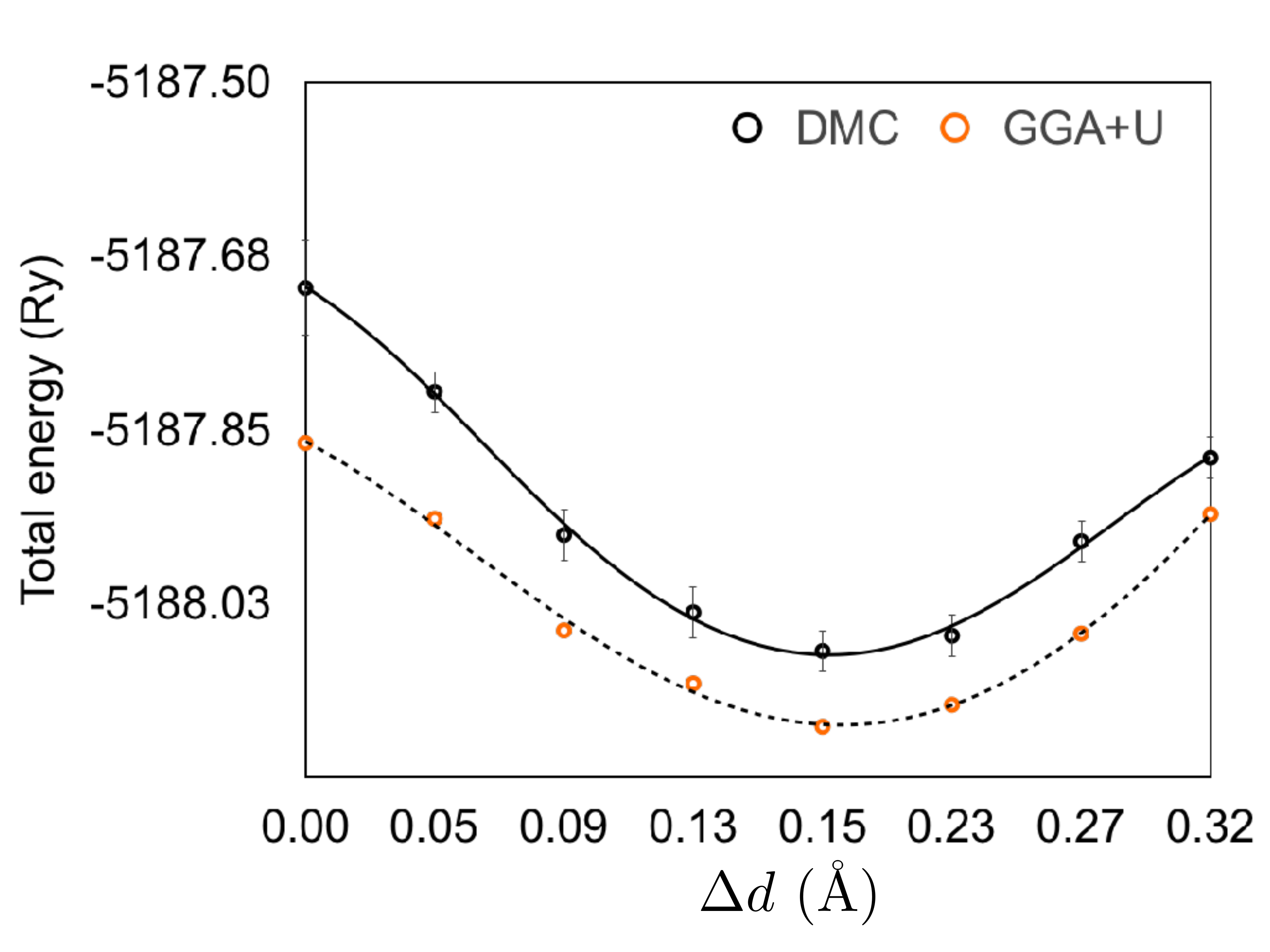}
\caption{(Color online) GGA+U and DMC total energies for the $V_{\rm O}^{+2}$ defect state as a function of parameter $\Delta d$ in units of~\AA. The data were fitted with a polynomial function, shown as a solid line (GGA+U) and as a dashed line (DMC).} 
\label{relax}
\end{figure}

In all the following calculations, we used DFT-optimized structures. Ideally, we should use the DMC-optimized structures, but DMC calculations of interatomic forces for large systems are at the present impractical. However, in order to gauge the difference between DFT- and DMC-optimized structures, we investigated a test case with a single +2-charged oxygen vacancy ($V_O^{+2}$). In cubic hafnia, oxygen atoms occupy all tetrahedral sites, and the  hafnium atom forms a face-centered cubic lattice. Near the vacancy site, the \nth{1} nearest neighbor (NN) shell is occupied with four Hf atoms and the \nth{2} NN shell is filled with six oxygen atoms. 
From the DFT calculations, we found a spherically symmetric  relaxation of the ions on \nth{1} NN shell of the vacancy site. To study the relaxation of ions near the vacancy site at the DMC level we considered a parameter, $\Delta d$, which is the spherically uniform relaxation of the \nth{1} NN shell relative to the vacancy site. For example, $\Delta d=0.0$ corresponds to an un-relaxed systems with a $V_{\rm O}^{+2}$ oxygen vacancy. With the NN shell displaced by $\Delta d$, the ions in the \nth{2} NN shell were displaced while maintaining a constant ratio between the nearest and next-nearest shell distances to the defect site. The total energies for GGA+U and DMC for different values of $\Delta d$ values for the $V^{\rm +2}_{\rm O}$ state are shown in Fig.~\ref{relax}. 
The total energy minima were at $\Delta d=0.176(2)$~{\AA} and $\Delta d=0.185$~{\AA} for DMC and GGA+U, respectively. This indicates that using the DFT relaxed structures introduces only minor errors at the DMC level. 
\subsection{Oxygen vacancies}
We start our discussion with neutral oxygen vacancies. In the DFT structural relaxation, the adjacent Hf ions were  relaxed towards the vacancy position by a displacement of 0.02~\AA~from their ideal positions, which corresponds to 0.89\% of the Hf-O bond length. The relaxation of Hf atoms is driven by the electronic configuration around the vacancy site: removing an oxygen atom from a perfect cubic hafnia crystal leaves  two extra electrons, with each Hf dangling bond contributing 1/2 electron in its $5d$ shell. The total energy of the system is then minimized by slightly contracting the Hf atoms towards the vacancy site. However, in the positively charged defect states V$_{\rm O}^{\rm +1}$ and V$_{\rm O}^{\rm +2}$, the adjacent Hf ions relax 0.084~{\AA} and 0.180~\AA, respectively, outward from the vacancy site, consistent with the findings in Ref.~[\onlinecite{XIONG2005408}], driven by the repulsion between the positively charge Hf ions and the positively charged vacancy. In contrast, for negatively charged defects V$_{\rm O}^{\rm -1}$ and V$_{\rm O}^{\rm -2}$, obtained by adding one and two electrons, respectively, to the neutral vacancy, the Hf ions relax inwards by about 0.03~{\AA}  and 0.13~\AA, respectively, because of the Coulombic attraction to the vacancy site.
\subsubsection{Defect formation energies}
\begin{table*}[!ht]
\centering
   %  \small\addtolength{\tabcolsep}{-0pt}
   \begin{large}
  \begin{tabular}{|l|l|l|l|l|}
    \hline
    \multirow{3}{*}{Charge} &
      \multicolumn{2}{c}{DFT} &
      \multicolumn{2}{c|}{DMC}   \\
      \cline{2-5}
    & Oxygen-rich & Oxygen-poor & Oxygen-rich & Oxygen-poor \\
    \hline
    -2 & 14.67 & 9.18 & 16.15(6) & 9.64(6)  \\
    \hline
     -1 & 10.49 & 5.00 & 11.72(5) & 5.20(5) \\
    \hline
     0 & 6.19 & 0.71 & 7.23(2) & 0.72(1)  \\
    \hline
     +1 & 3.12 & -2.37 & 4.25(4) & -2.26(4)  \\
    \hline
     +2 & -0.38 & -5.87 & 0.24(6) & -6.27(9)  \\
    \hline\hline
    N & 5.90 & 0.41 & 6.79(6) & 0.27(1)  \\
    \hline
  \end{tabular}
   \end{large}
  \caption{Defect formation energies in eV for different oxygen vacancy charge states, and for a neutral substitutional N defect.}
    \label{tab:1}
\end{table*}
The calculated formation energies of these states under oxygen-rich and oxygen-poor conditions are listed in Table~\ref{tab:1}. The formation energy of neutral and charged defects was calculated using 96-atom cells and including all previously described corrections (twist averaging, Chiesa and Model Periodic Coulomb interactions). The calculated GGA+U valence band minimum $E_{VBM}$ for different states is used for both GGA+U and DMC formation energies.
The calculated GGA+U and the DMC formation energies for a neutral vacancy state are, respectively, 6.19~eV and 7.23(12)~eV (oxygen-rich conditions), and 0.71~eV and 0.72(11)~eV (oxygen-poor condition). For a comparison, the GGA+U values are lower than the results from GGA calculations of J. X. Zheng {\it et al.}~\cite{PhysRevB.75.104112} who obtained a value of 6.63~eV (oxygen-rich condition) and 0.98~eV (oxygen-poor condition) for the fourfold coordinated oxygen in monoclinic HfO$_2$. Also, we note that the obtained formation energy for a neutral oxygen vacancy is smaller than a value of 6.95~eV obtained for cubic HfO$_2$ from GGA calculations~\cite{doi:10.1063/1.4989621}    The GGA+U and the DMC formation energies for V$^{\rm -1}_{\rm O}$ and V$^{\rm -2}_{\rm O}$ states are higher than the neutral oxygen vacancies, indicating that these vacancies are unstable when compared to the neutral vacancy.
Therefore, we do not discuss these charged defect states further. Both the GGA+U and the DMC formation energies for V$^{\rm +1}_{O}$ and V$^{\rm +2}_{O}$ are lower than the formation energy of neutral oxygen vacancy under oxygen-rich and oxygen-poor conditions, indicating that both these states are more stable than the neutral and negatively charged oxygen vacancies.
\subsubsection{Negative-U$^{\prime}$ effect}
To understand the disproportion of -2, -1, 0, +1 and +2 charge oxygen vacancies in hafnia, we computed an effective  $U^{\prime}$ parameter (which is different from the Hubbard U in GGA+U). The $U^{\prime}$ parameter has a physical meaning: it captures the quantitative repulsive electrostatic interaction ($U_{\rm el}$) between the ionic defects, and the electron-lattice relaxation energy ($U_{\rm rel}$),
\begin{enumerate}[label=(\roman{*})]
\item{Injecting an electron into hafnia, ${2V^{-1}_{\rm O} \rightarrow V^{0}_{\rm O}+V^{-2}_{\rm O}}$:}
When an electron is injected into the hafnia material, a neutral vacancy state traps it and becomes $V^{-1}_{\rm O}$, and adding an additional electron to a $V^{-1}$ state creates a $V^{\rm -2}_{\rm O}$ defect. The energetics of the reactions in terms of $U^{\prime}$ are obtained as $U^{\prime}=E[V^{0}_{\rm O}]+E[V^{-2}_{\rm O}]-2E[V^{\rm -1}_{\rm O}]$, where $E[\ldots]$ is the energy of system for a corresponding charge defect state.  While the computed GGA+U value for $V_{O}^{1}$  ($-0.12$~eV) suggests that the $V_{O}^{1}$ defect is unstable (exothermic process) when compared to $V_{O}^{0}$ and $V_{O}^{-2}$, the corresponding DMC value ($-0.06(8)$~eV) is not accurate enough to draw a similar conclusion.

\item{Injecting a hole into hafnia, ${2V^{\rm +1}_{\rm O} \rightarrow V^{\rm 0}_{\rm O}+V^{\rm +2}_{\rm O}}$:} When a hole is injected into hafnia, a neutral vacancy traps it and becomes a $V^{+1}_{\rm O}$ vacancy. By further adding a hole to the $V^{+1}_{O}$ state, a $V^{+2}_{\rm O}$ state is created. In this case, $U^\prime=E[V^{\rm 0}_{\rm O}]+E[V^{\rm +2}_{\rm O}]-2E[V^{\rm +1}_{\rm O}]$. The obtained $U^{\prime}$ for GGA+U (-0.43~eV) and DMC (-1.03(1)~eV) indicate that the $V^{+1}_{\rm O}$ is very unstable when compared to $V^{0}_{O}$ and $V^{+2}_{O}$.

\item{Neutral defect creation in hafnia, ${V^{-2}_{\rm O}+V^{+2}_{\rm O}\rightarrow V^{0}_{\rm O} }$:} This process occurs by de-trapping charges from the charged vacancies,  and $U^{\prime}=2E[V^{0}_{\rm O}]-E[V^{-2}_{\rm O}]-E[V^{+2}_{\rm O}]$. The calculated GGA+U (-1.91~eV) and DMC (-1.45(3)~eV) values indicate that the charge state $V^{\rm 0}_{\rm O}$ is more stable than $V^{\rm -2}_{\rm O}$ and $V^{\rm +2}_{\rm O}$. 
\end{enumerate}

In all these three cases, we obtain a negative $U^{\prime}$ value for both GGA+U and DMC. The calculated DMC $U^{\prime}$ value for $E[V^{+2}_{O}]$ is lower than the GGA+U one by about 0.6~eV. This seems to suggest a strong lattice relaxation at the vacancy site in a supercell at the GGA+U level,  compared to DMC. However, as shown in Fig.~\ref{relax}, we found that the relaxations in GGA+U and DMC yield almost the same location of the minimum displacements. The possible cause of this discrepancy may then lie in the repulsive electron interaction terms: the electronic correlations in GGA+U are considerably more approximate than in DMC, which may ultimately lead to a higher GGA+U $U^{\prime}$ value for the same displacements.

\subsubsection{Density of states (DOS)}
\begin{figure}[!ht]
\includegraphics[width=1.0\columnwidth]{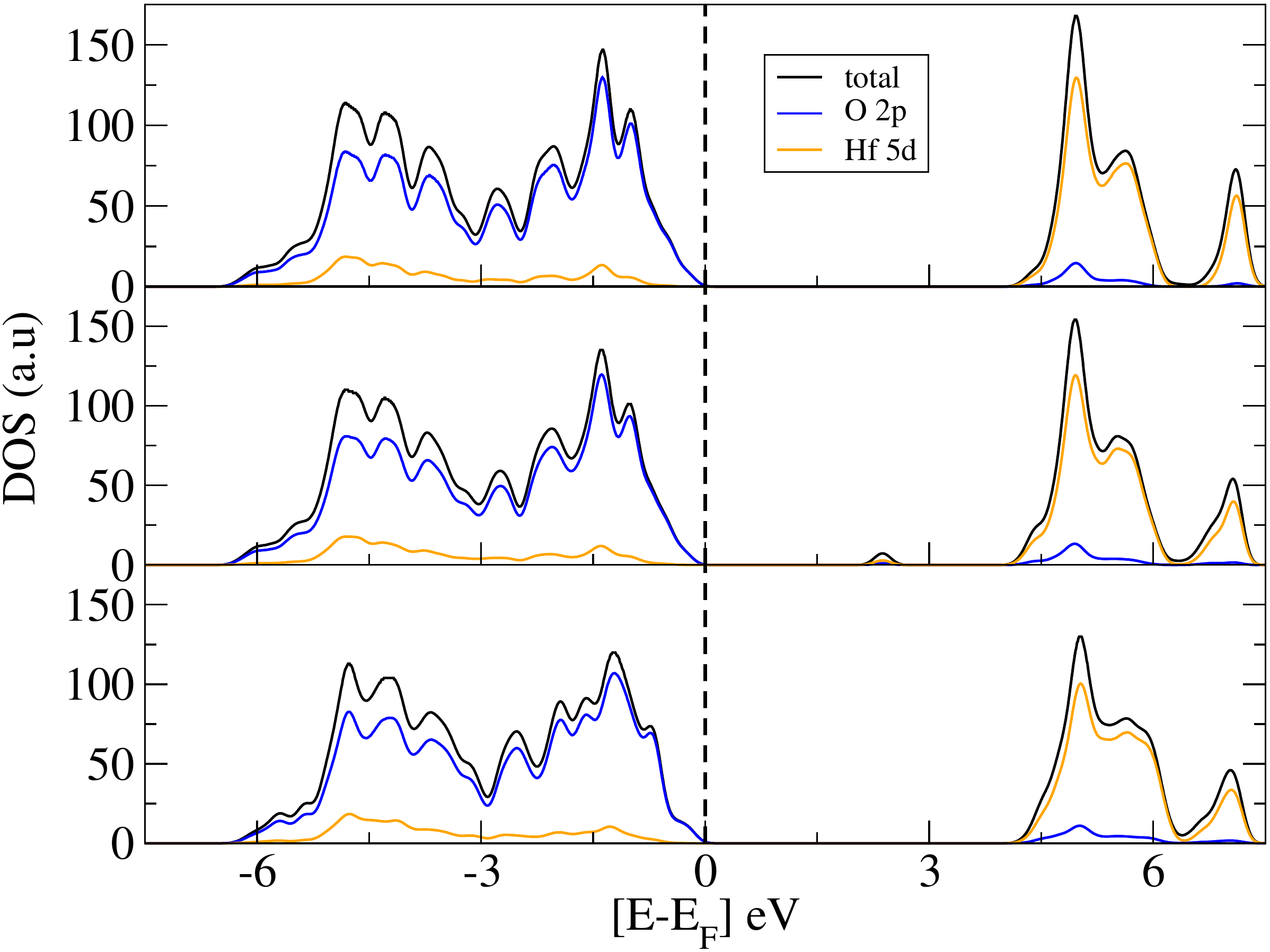}
\caption{(Color online) DOS and PDOS (O $2p$ and Hf $5d$) for stoichiometric cubic hafnia (top panel), cubic hafnia with a neutral O vacancy (center panel), and with a +2 O vacancy state (bottom panel). The Fermi level is set to the valence band edge (dashed line).} 
\label{fig:1}
\end{figure}

\begin{figure}[!ht]
\includegraphics[width=1.0\columnwidth]{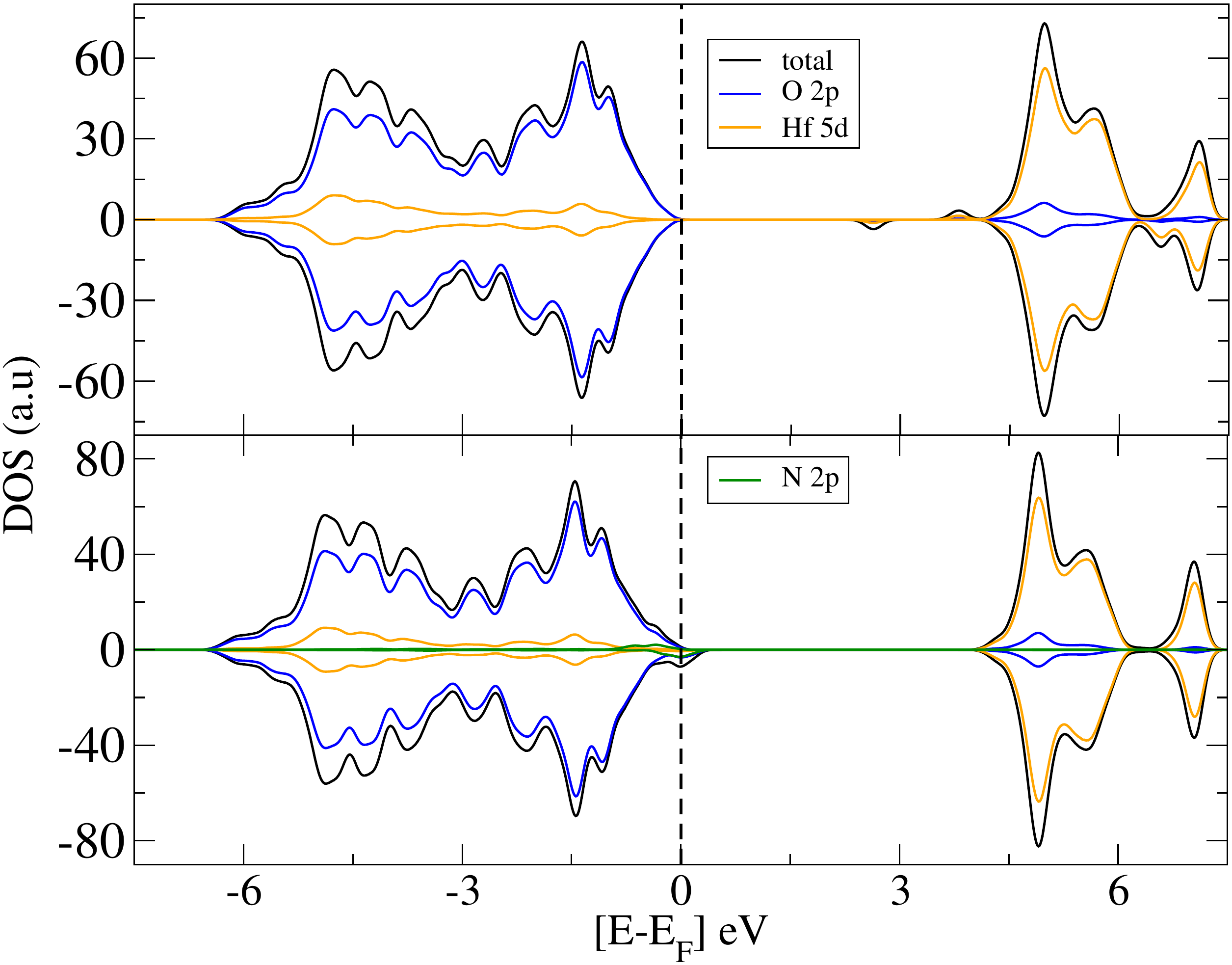}
\caption{(Color online) Total density of states for hafnia with a V$^{\rm +1}_{\rm O}$ defect state (top panel), and N substitutional defect (bottom panel). The magnetism in the N-doped system is mainly due to the N $2p$ at the Fermi level, while in the V$^{\rm +1}_{\rm O}$ it is due to the Hf $5d$ states. }
\label{fig:2}
\end{figure}

The GGA+U total electron density of states (DOS) and the orbital-projected density of states (PDOS) for pure, V$^{\rm 0}_{\rm O}$ and V$^{\rm +2}_{\rm O}$ charge oxygen vacancies of hafnia are shown in Fig.~\ref{fig:1}.
The calculated GGA+U band gap for hafnia is 4.04~eV, which is smaller than the experimental value of 5.8~eV\cite{doi:10.1063/1.1456246}. The PDOS shows that the oxygen $2p$ states dominate the top of the valence band, while hafnia $5d$ states contribute to the bottom of the conduction band. For a neutral oxygen vacancy, a defect state is created in the middle of the gap. The defect state is strongly localized on the $5d$ orbitals and $2p$ of the adjacent Hf ions and O ions, respectably. The mid-gap defect state leads to an effective reduction of the band gap to about $2.0$~eV. For the V$^{\rm +2}_{\rm O}$ charge state, the band gap is reduced to 3.9~eV, and there are no defect states in the gap. Figure~\ref{fig:2} shows DOS and PDOS for a charge +1 O vacancy (top panel). This defect state has an induced moments. We found a total magnetic moment of -1~$\mu_{B}$ per hole in V$^{\rm +1}_{\rm O}$ state
\subsection{Substitutional defects}
We have also investigated the effect of substitutional doping with nitrogen at an oxygen site in the supercell. This creates a substitutional N defect with a defect concentration of 1.04\%. 
The positive charge (one hole) at the N site leads to a small inward relaxation of the nearest hafnium shell, and an outward relaxation of neighboring oxygen atoms near the N site. These displacements are 0.004~{\AA} and 0.03~{\AA}, respectively. Also, we found that a spin-polarized structure  with a total magnetic moment of +1~$\mu_{B}$ is more stable than an un-polarized one. The total moment is mostly residing on the single $2p$ hole provided by the N dopant; N has a magnetic moment of $\sim 0.4~\mu_{B}$, and the near-neighbor Hf atoms and 2nd NN O attain induced ferromagnetic moments, with a magnetic moment of 0.0012~$\mu_{B}$ and 0.0486~$~\mu_{B}$, respectively. This is illustrated in the bottom panel of Fig.~\ref{fig:2}.
\begin{figure*}[htp]
\includegraphics[trim={0 5cm 0 5cm},clip,width=2.0\columnwidth]{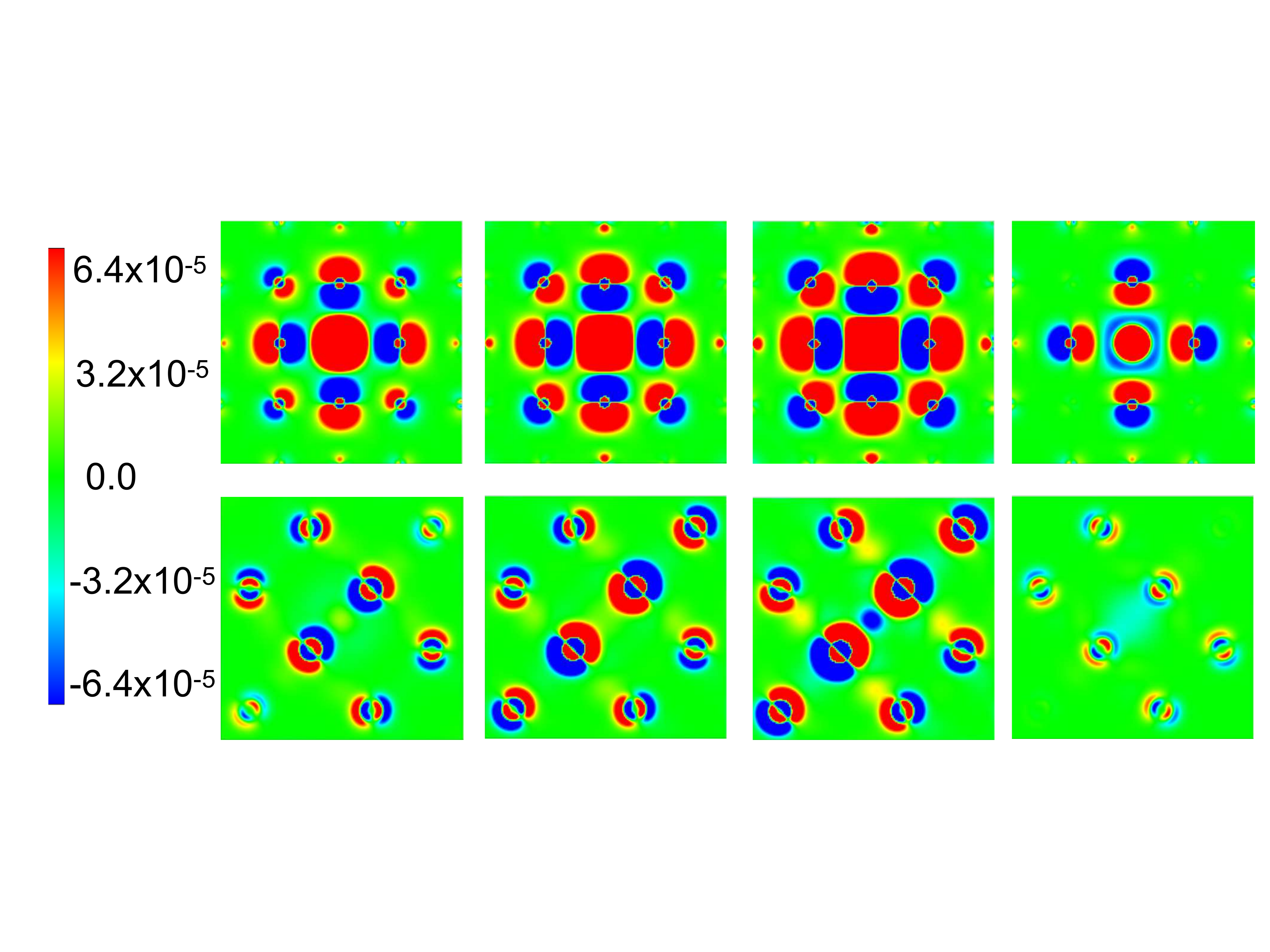}
\caption{(Color online) Two-dimensional contour plots of GGA+U total electron density differences between the pure system and the V$^{\rm 0}_{\rm O}$,  V$^{\rm +1}_{\rm O}$ , V$^{\rm +2}_{\rm O}$, and N-dopant systems, respectively. The top panels show the total electron density difference in the central (001) plane; the lower panels for the atomic plane below the central (001) plane. The color scale indicates number densities from $-6.4\times10^{-5}$ to $+6.4\times10^{-5}$~\AA$^{-3}$. Red (positive electron density) means that the defective system has smaller electron density than the ideal system, and blue indicates an excess electron density in the defective system relative to the ideal one. } 
\label{fig:3}
\end{figure*}
\begin{figure*}[htp]
\includegraphics[trim={0 5cm 0 5cm},clip,width=2.0\columnwidth]{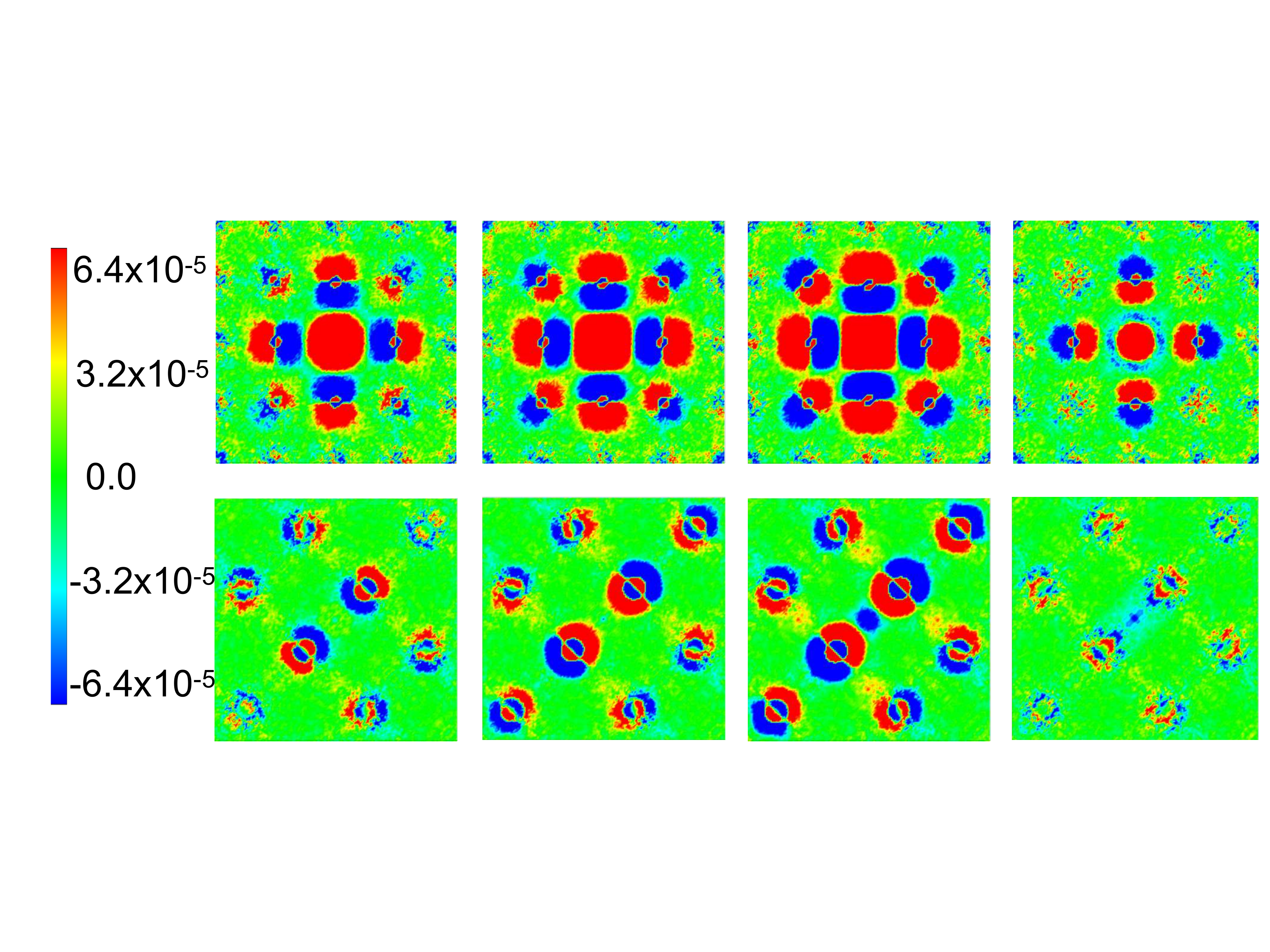}
\caption{(Color online) Two-dimensional contour plots of DMC difference total electron densities between the pure system and the V$^{\rm 0}_{\rm O}$,  V$^{\rm +1}_{\rm O}$ , V$^{\rm +2}_{\rm O}$, and N-dopant systems, respectively. The top panels show the total electron density difference in the central (001) plane; the bottom panels for the atomic plane below the central (001) plane. The color scale bar indicates number densities from $-6.4\times10^{-5}$ to $+6.4\times10^{-5}$~\AA$^{-3}$.} 
\label{fig:4}
\end{figure*}

\begin{figure}[!ht]
\includegraphics[trim={0 1cm 0 1cm},clip,width=1.0\columnwidth]{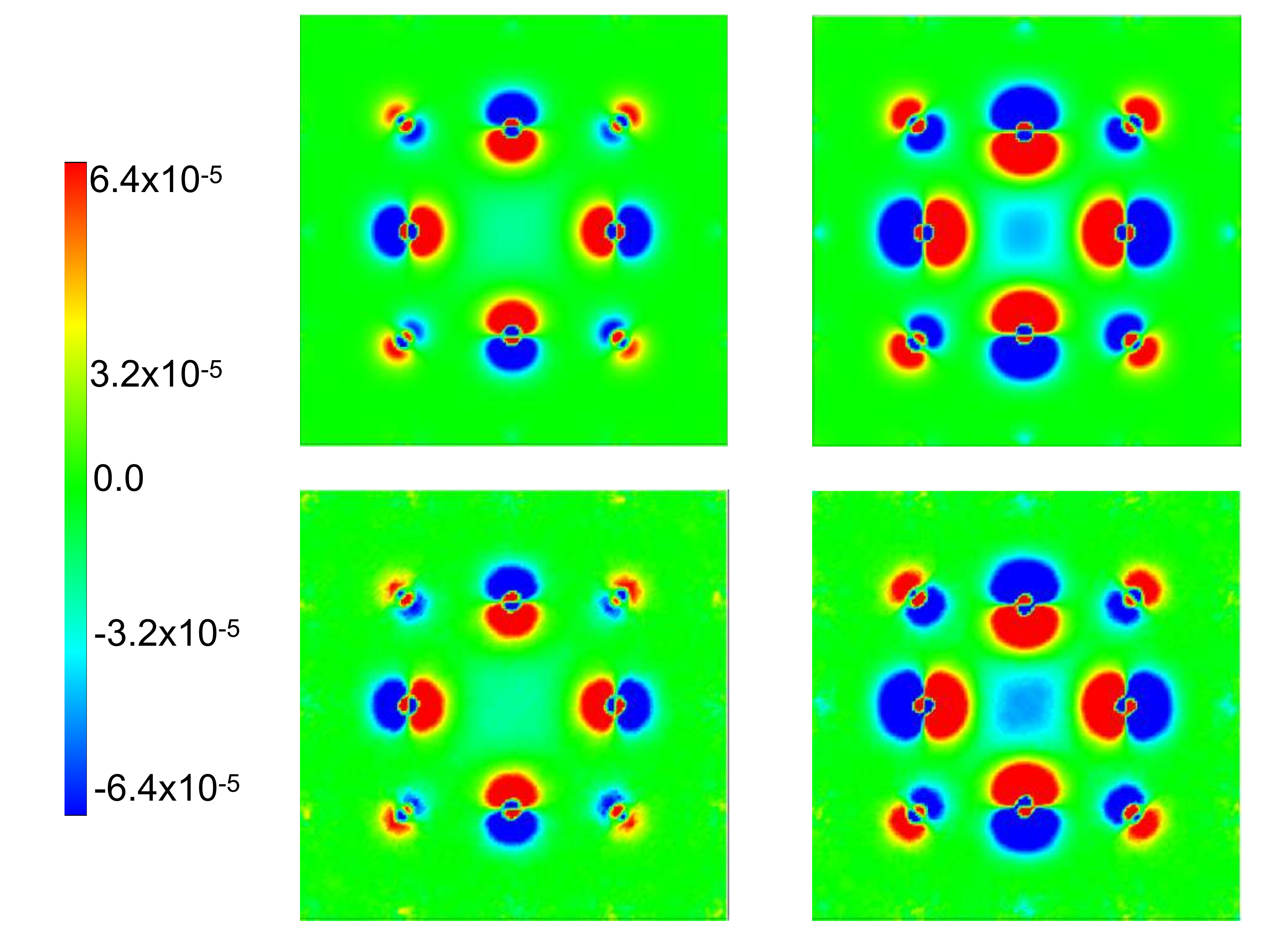}
\caption{(Color online) Two-dimensional contour plots of DFT (top panel) and DMC (bottom panel) total electron densities difference in the (001) basal plane between the   V$^{\rm 0}_{\rm O}$ system and the V$^{\rm +1}_{\rm O}$ , and V$^{\rm +2}_{\rm O}$ systems, respectively. The color scale goes from $-6.4\times10^{-5}$ to $6.4\times10^{-5}$~\AA$^{-3}$.} 
\label{fig:5_1}
\end{figure}
Close to the Fermi level, E$_{F}$, we note a significant hybridization between the N $2p$ and O $2p$ states. %. 
The hybridization leads to a splitting of energy levels near E$_F$. The spin-split states near E$_{F}$ result in a ferromagnetic insulator character, with a band gap of 3.6~eV. This band gap is about the same as the one in pure cubic hafnia, which suggests that by leakage currents can be reduced by nitrogen doping, for example by annealing in nitrogen-rich atmosphere, to eliminate oxygen vacancies with their mid-gap states.  
Earlier  first-principles calculations reported that the incorporation of two N atoms next to the oxygen vacancy sites shifts the vacancy level out of the gap~\cite{4160052,XIONG2005408} which is consistent with our results. Recently, measured current density versus voltage curves reported a decrease of leakage current density and a lower number of interface trap charges in pre-nitrated orthorhombic films~\cite{Maurya2018}. Our results indicate that the substitution of a single nitrogen atom at the oxygen vacancy site removes the single defect level created by the neutral oxygen vacancy which ultimately reduces the leakage currents. However, the excess charge at the N-dopant site still poses a fixed-charge problem which may be resolved by doping one more N atom in the supercell. This may be the subject of future studies. 

We have also computed the formation energy E$_f$(N) of an N-dopant in cubic hafnia by using Eq.~\ref{eq2} to assess the thermodynamic stability of the N-dopant under oxygen-rich conditions or oxygen-poor conditions. The required nitrogen chemical potential is obtained as   $\mu_{\rm N} \approx \frac{1}{2} {\rm E}({\rm N}_2)$, where $E(N_2$) is the total energy of a nitrogen dimer. The computed chemical potential of single nitrogen atoms is 271.44~eV (GGA+U) and 271.24(1)~eV (DMC). The formation energy of a substitution N dopant is listed in Table~\ref{tab:1}. The calculated formation energies from GGA+U and DMC, respectively, for an N-dopant are lower than the computed formation energies for neutral oxygen vacancies under both oxygen-rich conditions and oxygen-poor conditions. The GGA+U N-dopant formation energy is lower than the DMC formation energy, consistent with GGA+U underestimating the cohesive energy for hafnia compared to DMC\cite{PhysRevMaterials.2.075001}. 
This suggests that the formation of N substitutional defects is very likely under normal oxygen atmosphere conditions.

The computed DMC and DFT formation energies for different oxygen vacancy charge states and a neutral N dopant are presented in Table~\ref{tab:1}. In general, the DMC values under oxygen-rich conditions are higher by 0.6--1.5~eV than the corresponding GGA+U formation energies. We attribute the differences between the DMC and GGA+U energies to the different description of the $5d$-orbitals, and to the different description of the correlation energy, similar to observations of defects in $3d$ transition metal oxides: GGA+U typically underestimates the formation energies and band gap\cite{PhysRevMaterials.1.065408,PhysRevMaterials.1.073603}. 
Note, however, that while the DMC defect formation energy is typically larger than the GGA+U one, the DMC formation energy of a charge +2 oxygen vacancy and of neutral N dopant under oxygen-poor conditions are lower than the corresponding GGA+U energies by 0.4~eV and 0.14~eV, respectively. 
Given that experimental formation energy values for the cubic phase are lacking, it is our hope that our computed DMC values will serve as useful benchmarks.

\begin{figure}[!htp]
\centering
\includegraphics[trim={0 0cm 0 0cm},clip,width=1.0\columnwidth]{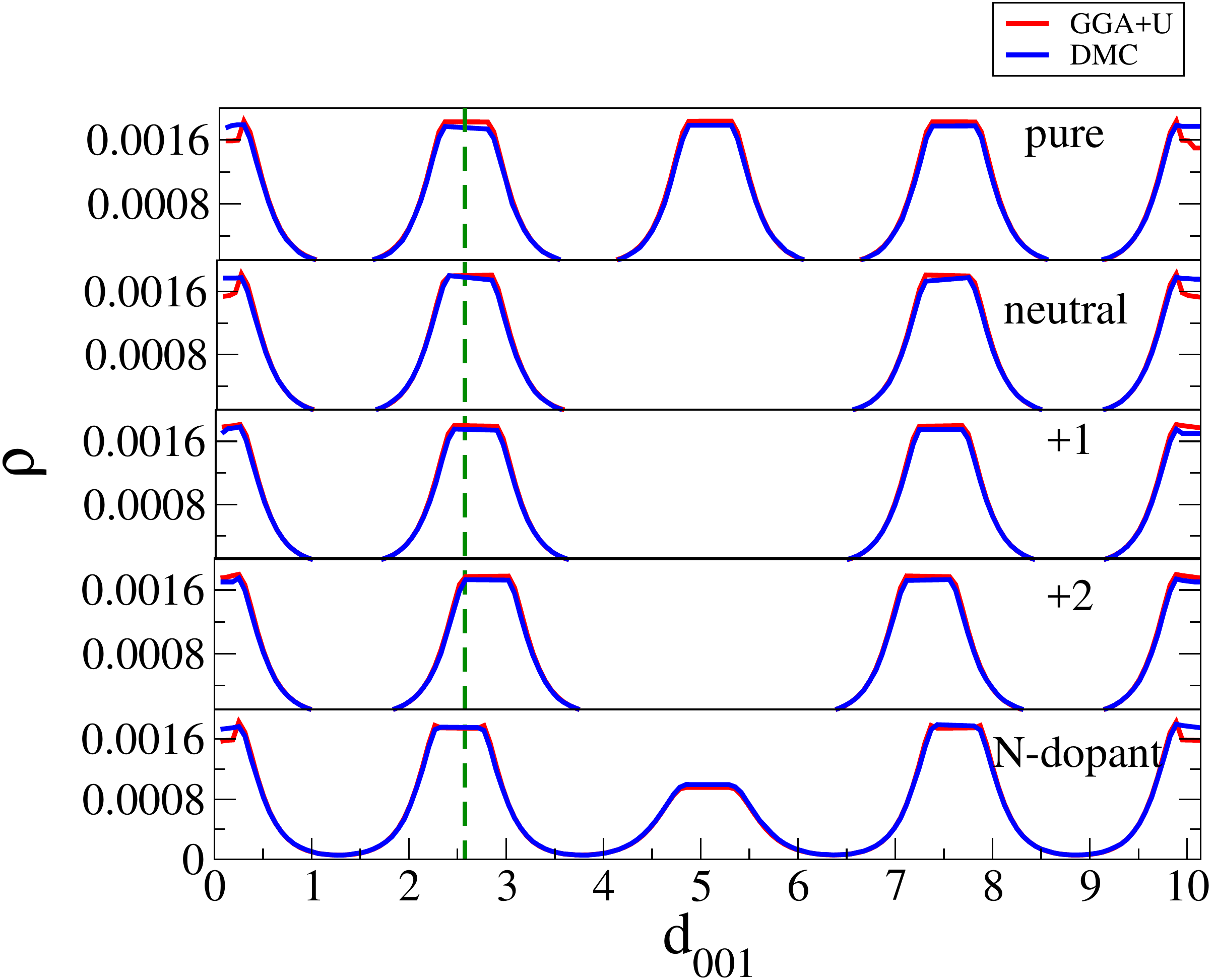}
\caption{(Color online) One-dimensional symmetrized charge density along the (001) direction through the point  $(5.07,5.07,z)$~{\AA} for a pure, neutral, V$^{\rm +1}_{\rm O}$, V$^{\rm +2}_{\rm O}$ and N-doped HfO$_2$ from top to bottom panels respectively.  The defect is centered at $(5.07,5.07,5.07)$~{\AA}. The line indicates the displacement of ions from their original position which leads to displacements of charge densities along the axis.} 
\label{1dplot}
\end{figure}
\begin{figure}[!ht]
\includegraphics[trim={0 1cm 0 1cm},clip,width=1.0\columnwidth]{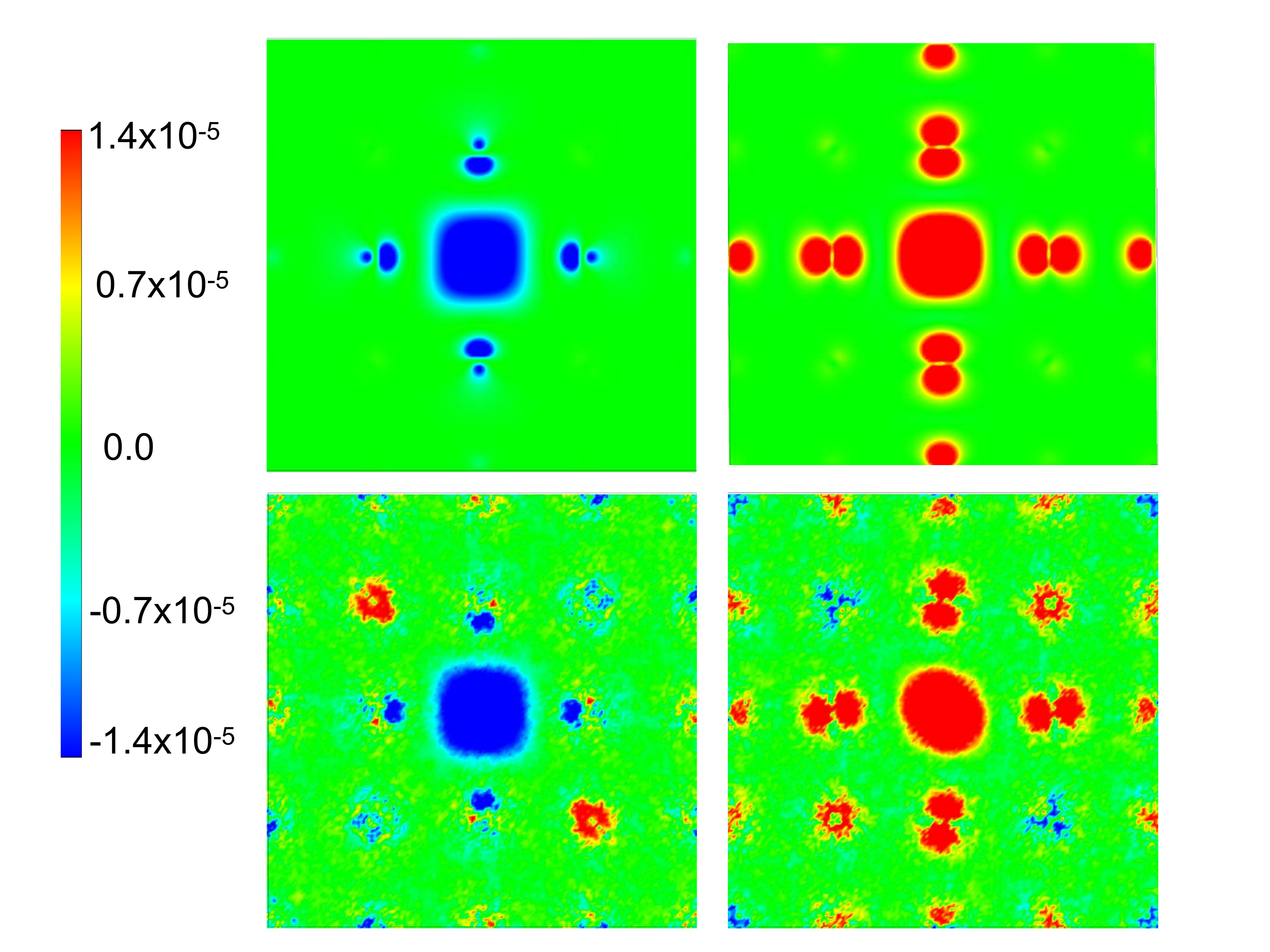}
\caption{(Color online) Two-dimensional contour plots of the GGA+U (top panel) and the DMC (bottom panel) spin densities. The left panels show the spin density in the central plane along (001) basal plane with a vacancy V$^{\rm +1}_{\rm O}$ state; the right panels show the spin density in the central (001) plane with the N dopant at the center of the plot. The color scale goes from $-1.4\times10^{-5}$ to $1.4\times10^{-5}$~\AA$^{-3}$. The designation of up- and down-spin (blue {\em vs.} red) is arbitrary and irrelevant. Spin densities are concentrated at center and oxygen sites for the V$^{\rm +1}_{\rm O}$ state, and at the center of the N site and on the \nth{2} NN oxygen sites for a N-dopant.} 
\label{fig:5}
\end{figure}
\subsection{Charge densities}
We calculated the total electron density distribution in supercells with V$^{\rm 0}_{\rm O}$, V$^{\rm +1}_{\rm O}$, V$^{\rm +2}_{\rm O}$ and a neutral N-dopant in order to analyze differences between the GGA+U and the DMC electron densities and spin densities. The calculated raw DMC data are noisier because of the statistical sampling, so we reduced the noise by averaging. For a center plane through the vacancy (oxygen plane) and below the center plane (hafnium plane), we averaged the charge or spin density by a 180-degree rotation about a (001) axis, and by reflections in (110) and ($\bar{1}$10) planes.  
In general, the GGA+U and DMC charge densities are qualitatively and quantitatively rather similar with only some minor quantitative differences that will be detailed below. The close similarities between GGA+U and DMC charge densities were also noted in previous work on NiO\cite{PhysRevMaterials.1.073603}.

Figure~\ref{fig:3} shows the differences in electron densities between the pure and the V$^{\rm 0}_{\rm O}$, V$^{\rm +1}_{\rm O}$, V$^{\rm +2}_{\rm O}$, and single neutral N-dopant systems obtained using GGA+U, and the corresponding (symmetrized) DMC electron density differences are shown in Fig.~\ref{fig:4}. The top panels show the difference densities on a central plane (oxygen plane) along the (001) direction through the vacancy site, and the bottom panels show the difference densities in an atomic (001) plane just below the central plane. In going from the left-most panels in the two rows to the third panels from the left, the charge state of the vacancy increases from 0 to $+2$. That is reflected in the excess electron density in the center of the panels in the top row. The large dumbell-shaped electron distributions along the (100) and (010) directions are the \nth{2} NN oxygen shell and illustrate the inward distortion of the oxygen atoms and their electron distributions towards the vacancy; the smaller dipolar distributions are on the \nth{3} NN hafnium atoms that are slightly distorted outward from the vacancy. Similarly, in the lower row of Figs.~\ref{fig:3} and \ref{fig:4} the two larger electron density distributions near the center of the panels are the \nth{1} NN hafnium atoms, that are displaced inward toward the neutral vacancy, and outward from the positively charged vacancies. In contrast, the right-most panels of Figs.~\ref{fig:3} and \ref{fig:4} show the outward displacement of the \nth{2} NN oxygens, and slight inward displacement of the \nth{1} NN hafnium atoms. 

In order to illustrate the hole concentration in the positively charged oxygen vacancy systems, we show in Fig.~\ref{fig:5_1} difference in GGA+U (top row) and DMC (bottom row) electron density between the V$_{\rm O}^{+1}$ and V$_{\rm O}$ systems, (left panels), and the V$_{\rm O}^{+2}$  and the V$_{\rm O}$ vacancy systems. The hole density in the V$_{\rm O}^{+1}$ and V$_{\rm O}^{+2}$ systems is clearly localized at the vacancy site, while the dipolar distributions just illustrated the further displacements of the \nth{2} oxygen ions towards the charged vacancies, and of the \nth{3} NN hafnium atoms away from the vacancy. Note that the DMC hole density difference at the vacancy site (bottom left panel in Fig.~\ref{fig:5_1}) appears to break the four-fold symmetry in that plane.

What is perhaps striking in comparing Figs.~\ref{fig:3}, \ref{fig:4}, and \ref{fig:5_1} is the good qualitative and quantitative agreement, at least at scale level of the figures, between the GGA+U and DMC charge densities. Indeed, only minor differences are discernible, such as the apparent larger density differences in the N-doped system on hafnium sites in DMC than in GGA+U, and the broken symmetry in the DMC V$_{\rm O}^{+2}$ hole density. This agreement is further illustrated in Fig.~\ref{1dplot}, which shows symmetrized GGA+U and DMC charge densities along the (001) direction through the center of the defect site at $(5.07,5.07,5.07)$~{\AA}. The figure clearly shows the atomic displacements, but also that the DMC charge density (blue line) is almost identical to the GGA+U one (red line).

\begin{figure}[!htp]
\centering
\includegraphics[trim={0 5cm 0 5cm},clip,width=1.0\columnwidth]{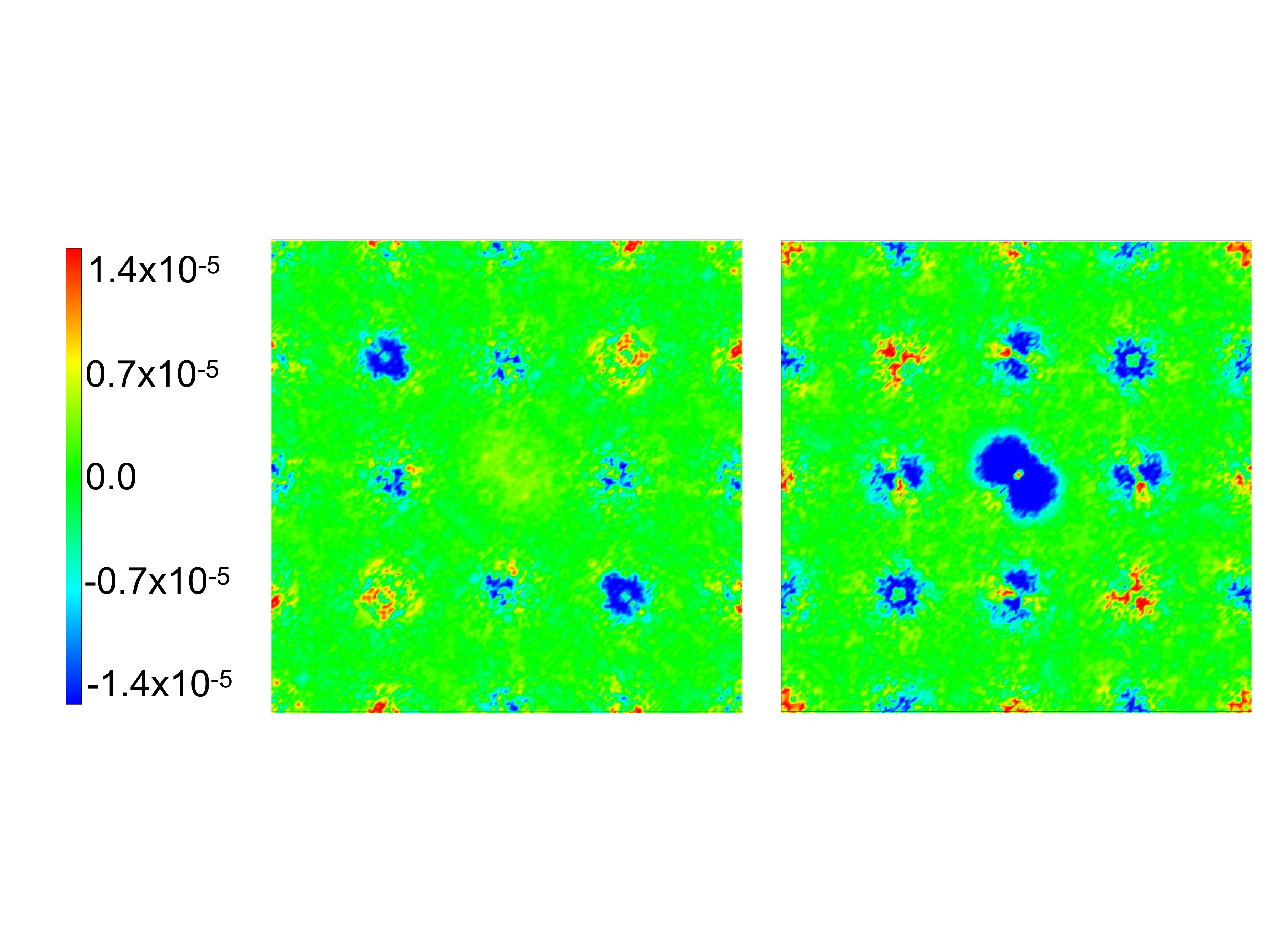}
\caption{(Color online) Two-dimensional contour plots of the spin density difference between GGA+U and DMC. The left panels show the spin density difference in the (001) basal plane with the vacancy V$^{\rm +1}_{\rm O}$ at the center; the right panels shows the spin density difference in the (001) basal plane with the N dopant at the center. The color scale goes from $-1.4\times10^{-5}$ to $1.4\times10^{-5}$~\AA$^{-3}$. Electrons are concentrated at Hf sites for the V$^{\rm +1}_{\rm O}$ vacancy and at the center for the N-dopant.} 
\label{fig:6}
\end{figure}
\begin{table}[!ht]
\centering
   %  \small\addtolength{\tabcolsep}{-0pt}
    \begin{tabular}{|P{3.5cm}|P{1.0cm}|P{1.0cm}|P{1.0cm}|P{1.0cm}|P{1.0cm}|}
     \hline
       Method &+1&+2&neutral&N&pure \\
       \hline
        RMSD($\rho_{\rm charge}$)(10$^{-5}$)&3.36&3.36&3.36&3.37&3.40 \\ 
        RMSD($\rho^{{\rm Hf(NN)}}_{\rm charge}$)(10$^{-5}$)&5.72  & 5.72& 5.71 & 5.72& 5.75  \\ 
        RMSD($\rho^{{\rm O(NN)}}_{\rm charge}$)(10$^{-5}$)& 5.05 &5.05 &5.72  &5.72 &  5.10 \\ 
        RMSD($\rho^{{\rm Hf}}_{\rm charge}$)(10$^{-5}$)& 5.72 & 5.72& 5.07 &5.07 &  5.75 \\ 
        RMSD($\rho^{{\rm O}}_{\rm charge}$)(10$^{-5}$)& 5.21 & 5.21& 5.54 &5.20 &  5.25 \\ 
        RMSD($\rho^{{\rm N}}_{\rm charge}$)(10$^{-5}$)& $-$ &$-$ & $-$ & 2.29 & $-$ \\
        RMSD($\rho_{\rm spin}$)(10$^{-5}$)& 0.24 &$-$ &$-$ &  0.29 & $-$ \\ 
        RMSD($\rho^{{\rm Hf(NN)}}_{\rm spin}$)(10$^{-5}$)& 0.24 &$-$ & $-$ &0.42 & $-$  \\ 
        RMSD($\rho^{{\rm O(NN)}}_{\rm spin}$)(10$^{-5}$)& 0.35 &$-$ & $-$ &0.44 & $-$  \\ 
        RMSD($\rho^{{\rm Hf}}_{\rm spin}$)(10$^{-5}$)& 0.36 &$-$ & $-$ &0.40 &  $-$ \\ 
        RMSD($\rho^{{\rm O}}_{\rm spin}$)(10$^{-5}$)& 0.35 &$-$ &$-$  & 0.42& $-$  \\ 
        RMSD($\rho^{{\rm N}}_{\rm spin}$)(10$^{-5}$)& $-$ &$-$ & $-$ & 1.60& $-$ \\ 
     \hline
    \end{tabular}
\caption{Charge and spin density RMSD for different oxygen vacancy charge states, and for an N-dopant. Statistical errors in the RMSDs are below 10$^{-8}$}
    \label{tab:2}
\end{table}
\subsection{Spin densities}
\begin{figure}[!htp]
\includegraphics[trim={0 5cm 0 5cm},width=1.05\columnwidth]{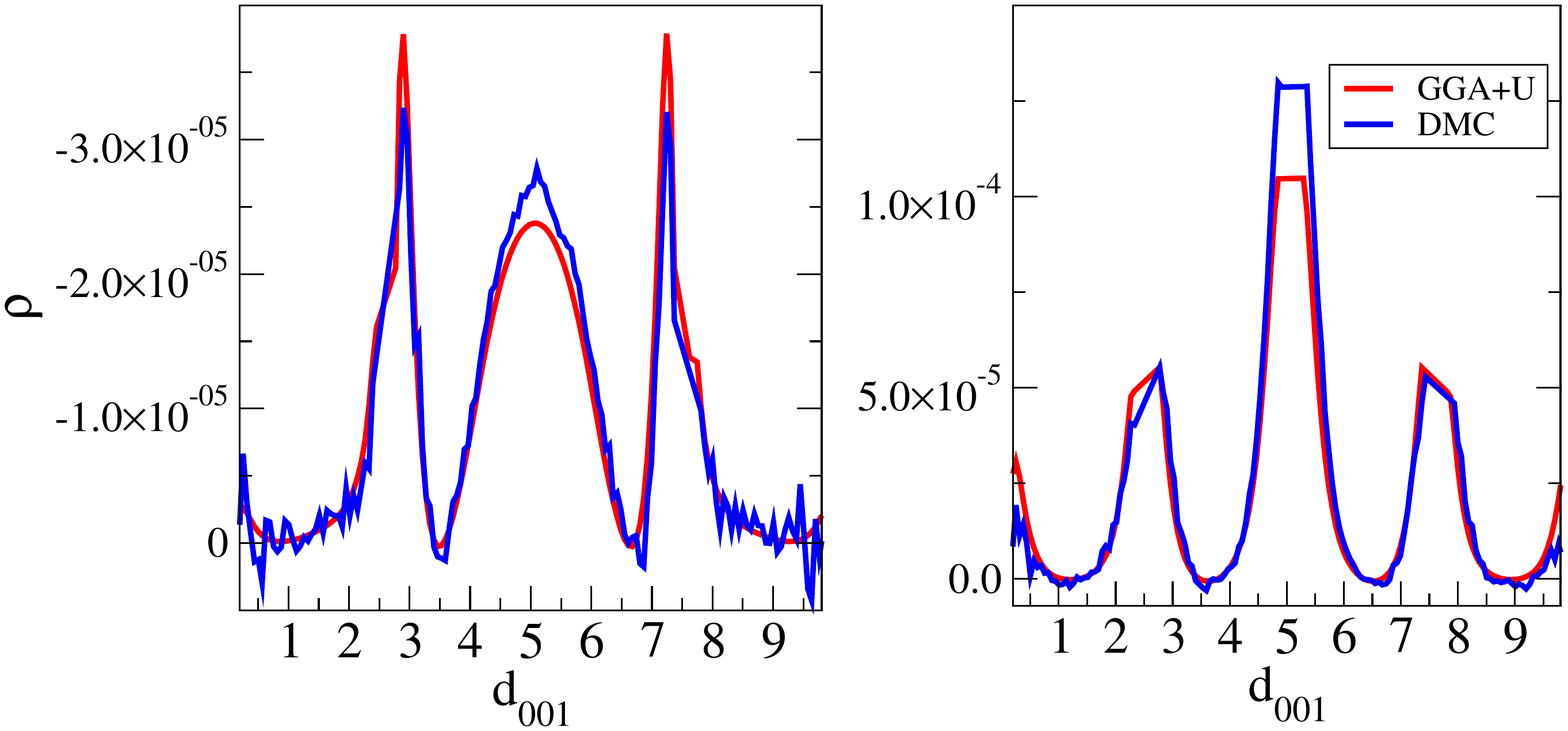}
\caption{(Color online) One-dimensional symmetrized spin density difference between GGA+U and DMC spin densities along the (001) direction through the point $(5.07,5.07,z)$~{\AA} for V$^{\rm +1}_{\rm O}$(left panel) and N-doped HfO$_2$ (right panel). The defect site is at center at $(5.07,5.07,5.07)$~{\AA}. 
The figures show a difference in spin density primarily on the defect site, with some difference also discernible on the \nth{1} NN Hf atoms.}
\label{fig:9}
\end{figure}
Fig.~\ref{fig:5} shows contour plots of the spin density for the V$^{\rm +1}_{\rm O}$ (left) and N-dopant (right) systems, calculated with GGA+U (top row) and DMC (bottom row), respectively, on a (001) plane through the center defect. In contrast with the electron densities in Figs.~\ref{fig:3} and \ref{fig:4}, the spin densities show some clear difference between GGA+U and DMC. First, DMC indicates a spin polarization on the \nth{3} NN hafnium atoms with opposite signs in the $(110)$ and $(\bar1,1,0)$ directions for both the $+1$ charged oxygen vacancy state and the nitrogen substitution state. Second, DMC spontaneously breaks the reflection symmetry of the spin density in $(100)$ and $(010)$ lines, which is clearly preserved by GGA+U. These differences are further highlighted in Fig.~\ref{fig:6}, which shows the spin density difference between the GGA+U and DMC spin densities for the V$_{\rm O}^{+1}$ system and the N-dopant system. Figure~\ref{fig:9} shows the one-dimensional symmetrized spin density difference between GGA+U and DMC along the (001) axis and through the center of the $V_{\rm O}^{+1}$ defect (left panel) and the N dopant (right panel). The main difference in both cases is in the spin density on the defect site, but with 

\subsection{Quantitative differences between GGA+U and DMC charge and spin densities}
In order to assess quantitative differences between the GGA+U and DMC charge and spin densities, we computed a root-mean-square deviation (RMSD) of the charge and spin densities as follows\cite{PhysRevMaterials.1.073603},
\begin{equation}\label{eq:RMSD}
%\begin{split}
{\rm RMSD}(\rho)=\sqrt[]{\frac{\sum_{i}^{\rm N}(\rho_{\rm DFT}({\rm R}_{i})-\rho_{\rm DMC}({\rm R}_{i}))^2}{\rm N}}
%\end{split} 
\end{equation}
$\rho_{\rm DFT}({\rm R}_{i})$ and $\rho_{\rm DMC}({\rm R}_{i})$ are the DFT and DMC charge/spin densities, respectively, and $i$ labels the $N$ gridpoints ${\rm R}_{i}$. The calculated RMSD($\rho_{\rm charge}$) and RMSD($\rho_{\rm spin}$) for different charges, pure and N-doped hafnia are listed in Table~\ref{tab:2}. To compute the RMSD near the Hf and O sites, we considered a spherical volume of grid points with a radius equal to half the bond length of Hf-O($\sim$1.0~\AA). The charge (spin) densities for Hf and O sites are $\rho^{\rm Hf}_{\rm charge}$ ($\rho^{\rm Hf}_{\rm spin}$) and $\rho^{\rm O}$ ($\rho^{\rm O}_{\rm spin}$), respectively. In addition, for the nearest neighbor Hf and O sites, the charge (spin) near to the vacancy site are $\rho^{\rm Hf(\rm NN)}_{\rm charge}$ ($\rho^{\rm Hf(\rm NN)}_{\rm spin}$) and $\rho^{\rm O(\rm NN)}_{\rm charge}$ ($\rho^{\rm O(\rm NN)}_{\rm spin}$), respectively. These RMSD values allow us to make some statements about differences in charge and spin distributions between DMC and GGA+U, but also about how charge redistributions in going from the pure to defective system differ between DMC and GGA+U.
\begin{enumerate}[label=(\alph{*})]
\item{Pure, V$_{\rm O}$, and N-dopant RMSD values:}
The pure, V$_{\rm O}$ and N-dopant systems show insignificant differences in the total charge RMSD and \nth{1} NN Hf charge RMSD: there is a relatively large difference in the charge distributions between GGA+U and DMC in the whole supercell and on the \nth{1} NN Hf atoms for each of these two systems, but these differences do not change between the pure and the V$_{\rm O}$ and N-dopant systems. However, the \nth{2} NN oxygen show a clear increase in the RMSD charge in going from the pure system to neutral oxygen vacancy, as do the hafnium and the oxygen charge RMSD values. This indicates that there is a charge redistribution on the \nth{2} NN oxygen shell that is different in DMC compared to GGA+U, and a charge redistribution on hafnium atoms {\em farther away} from the vacancy than the \nth{1} NN hafnium shell that is different in DMC compared to GGA+U. Note, however, that RMSD$(\rho^{\rm O}_{\rm charge})$ changes less in going from the pure system to the N-dopant, than from the pure system to the $V_{\rm O}$ system; there is less difference between the DMC and GGA+U charge redistribution on the oxygen farther away from the dopant in the N-dopant system than in the V$_{\rm O}$ system.
\item{Pure, and V$_{\rm O}^{+1}$, V$_{\rm O}^{+2}$ RMSD values:}
The computed charge RMSD values for the V$_{\rm O}^{+1}$ and the V$_{\rm O}^{+2}$ systems shows insignificant changes when compared to the pure system; there is no further differences between DMC and GGA+U charge redistributions compared for the positively charged vacancies compared to the pure system.
\item{Spin density RMSD for V$_{\rm}^{+1}$ and N-dopant systems:}
The $\rho^{\rm N}_{\rm spin}$ RMSD (1.60) is relatively large (compared to the other spin RMSD values) and similar to the $\rho^{\rm N}_{\rm charge}$ RMSD (2.29). This indicates that DMC and GGA+U differ both in charge and spin densities on the N dopant, as is illustrated for the spin density in Fig.~\ref{fig:6}. On the other hand, the other spin RMSD values are relatively small, consistent with the small but clearly discernible spin differences illustrated in Figs.~\ref{fig:6} and \ref{fig:9}.
\end{enumerate}
\section{Summary and Conclusions}\label{conclusion}
We have studied the oxygen vacancy defect states, and substitutional N-doping of cubic hafnia at absolute zero temperature using DFT and highly accurate DMC calculations. Because the cubic phase can be synthesized and stabilized in thin films at room temperature, we do not expect finite-temperature effects (ignored here) to be significant so long as the metastable free energy minimum is deep compared to thermal fluctuations at room temperature ($\sim25$~meV). In general, these would lead to slightly lower bulk modulus, which is not important for our study. For a longer discussion of finite-temperature effects on hafnia and zirconia polymorphs, please see Ref.~\onlinecite{PhysRevMaterials.2.075001}. Our calculations demonstrate that a substitutional N-dopant has much lower creation energy than that of a neutral oxygen vacancy under oxygen-poor conditions. Furthermore, the N-dopant does not introduce any mid-gap states that effectively lower the band gap. These results are consistent with the findings that nitrogen can passivate HfO$_2$\cite{Maurya2018}. Interestingly, the N dopant leads to a ferromagnetic state with about 0.4~$\mu_B$ per nitrogen. The DFT GGA+U and DMC defect formation energies are in reasonably good agreement with one another, especially for the formation energies under oxygen-poor conditions. The difference is larger for the formation energies under oxygen-rich conditions, but much of that is attributed to the difference in the DFT and DMC oxygen chemical potentials. We note that we checked that GGA+U and DMC give energy minima at the same structural distortions for the V$_{\rm O}^{+2}$ state, which gives us confidence that there are no significant discrepancies that may arise because the minimum-energy structures may be different. 

The positively charged oxygen defects, V$_{\rm O}^{+1}$ and V$_{\rm O}^{+2}$, have negative formation energies under oxygen-poor conditions, indicating that these defects will form spontaneously. However, stability analyses indicate that V$_{\rm O}^{+1}$ is unstable with respect to formation of neutral vacancies and V$_{\rm O}^{+2}$, and that the neutral vacancy V$_{\rm O}$ is stable with respect to formation of V$_{\rm O}^{+2}$ and V$_{\rm O}^{-2}$. Because the formation energy of V$_{\rm O}^{+2}$ and that of V$_{\rm O}^{+1}$ are negative under oxygen-poor conditions, such defects are expected to occur. It is therefore important to prevent charging (allowing electrons to escape) during formation of cubic hafnia to eliminate the creation of these charged oxygen vacancies; neutral defects can be eliminated with nitrogen passivation.

We also compared the GGA+U and DMC electron and spin densities. For the charge densities, the agreement is again reasonably good, but we note that DMC tends to break the four-fold symmetry in the (001) oxygen plane in the presence of charged oxygen vacancies. The spin densities of the magnetic V$_{\rm O}^{+1}$ and the N dopant show some larger qualitative differences in that DMC tends to polarize \nth{3} NN Hf antiferromagnetically, while GGA+U shows no discernible spin densities on the hafnium sites. Also, the DMC spin density of the N-dopant state shows a clear breaking of the four-fold symmetry in the (001) oxygen plane. 

Our work shows that for the structural properties and defect formation energies studies reported here, GGA+U is in reasonably good agreement with the much more accurate (and expensive) QMC calculations. The charge and densities are also in good qualitative agreement. There are, however, some important differences. Notably, DMC tends to yield somewhat more diffuse $d$-orbital that are slightly less localized than the GGA+U ones, and oxygen 2$p$ orbitals with a larger susceptibility to spin-polarization than the GGA+U ones (see, for example, Fig.~\ref{fig:5}).\cite{PhysRevMaterials.1.073603,song2018benchmarks} More importantly, in some cases, there is a startling {\em qualitative} difference in that QMC breaks the point group symmetry in the spin density about a defect site (Fig.~\ref{fig:5}). A detailed analysis reveals that such symmetry-breaking arises from the electronic correlations included accurately in QMC but only approximately, and in a rather uncontrolled approximation, in GGA+U. These correlations emanate from the 5$d$ orbitals in Hf. In addition, other electronic properties, such as the band gap, are not accurately captured by our GGA+U as the U-parameter was not specifically tuned to reproduce the experimental band gap. We conclude that while DFT schemes such as GGA+U do represent some physical features rather well both qualitatively and quantitatively, other features may not be even qualitatively accurately captured by DFT. The problem in applying DFT to systems with appreciable electronic correlations is that there is little systematic guidance to {\em a priori} assessments of what may be qualitatively inaccurate. It is our hope that studies such as ours can help guide DFT studies as well as improvements in DFT methods by providing highly accurate results.

\begin{acknowledgments}
 This research was carried out using resources of the Argonne Leadership Computing Facility, which is a DOE Office of Science User Facility supported under Contract DE-AC02-06CH11357. RC was supported by the Office of Science, U.S. Department of Energy, under Contract DE-AC02-06CH11357 as part of the Argonne Leadership Computing Facility Aurora Early Science Project. 
 HS, AB, and OH are supported by the U.S. Department of Energy, Office of Science, Basic Energy Sciences, Materials Sciences, and Engineering Division, as part of the Computational Materials 
Sciences Program and Center for Predictive Simulation of Functional Materials. The authors wish to thank J.T. Krogel and I. Kyl\"anp\"a\"a for helpful discussions and comments.  We gratefully acknowledge the computing resources provided on Bebop, high-performance computing clusters operated by the Laboratory Computing Resource Center at Argonne National Laboratory. 
\end{acknowledgments}
\appendix
\section{optimal parameter U}\label{sec:appen}

The remaining uncontrolled approximation in DMC is the position of the nodes from the initial trial wavefunction. Because VMC and DMC satisfy a strict variational principle, the nodal surface can be optimized in some parameter space by finding the minimum DMC energy in that parameter space. Following earlier work\cite{PhysRevMaterials.1.073603,PhysRevMaterials.2.075001,PhysRevMaterials.1.065408} we used DFT+U calculations with U as a parameter, and found the minimum DMC energy for the DFT+U trial wavefunctions as a function of U. We first performed self-consistent DFT GGA+U calculations using a 36-atom cubic hafnia supercell (twelve formula units) with a fixed lattice parameter of 5.12{\AA}~\cite{doi:10.1002/9780470318782.ch13}, a $6\times6\times6$ k-point mesh and a kinetic energy cut-off of 450~Ry. The GGA+U trial wavefunctions were then used in DMC calculations using the same supercell and with 64 twists. As shown in a previous QMC study of HfO$_2$\cite{PhysRevMaterials.2.075001},  there is a very small DMC energy difference of at most 0.02(4) eV/f.u. between the PBE+U and PBE trial wavefunctions. In order to estimate the optimal value of U, we interpolated the energies using a fourth order polynomial fit. The obtained optimal value from the fit was U$_{\rm opt}$ = 2.2(1)~eV with an excellent correlation coefficient of R$^{2}$=0.99.   

\begin{figure}%[!htp]
\includegraphics[width=1.00\columnwidth]{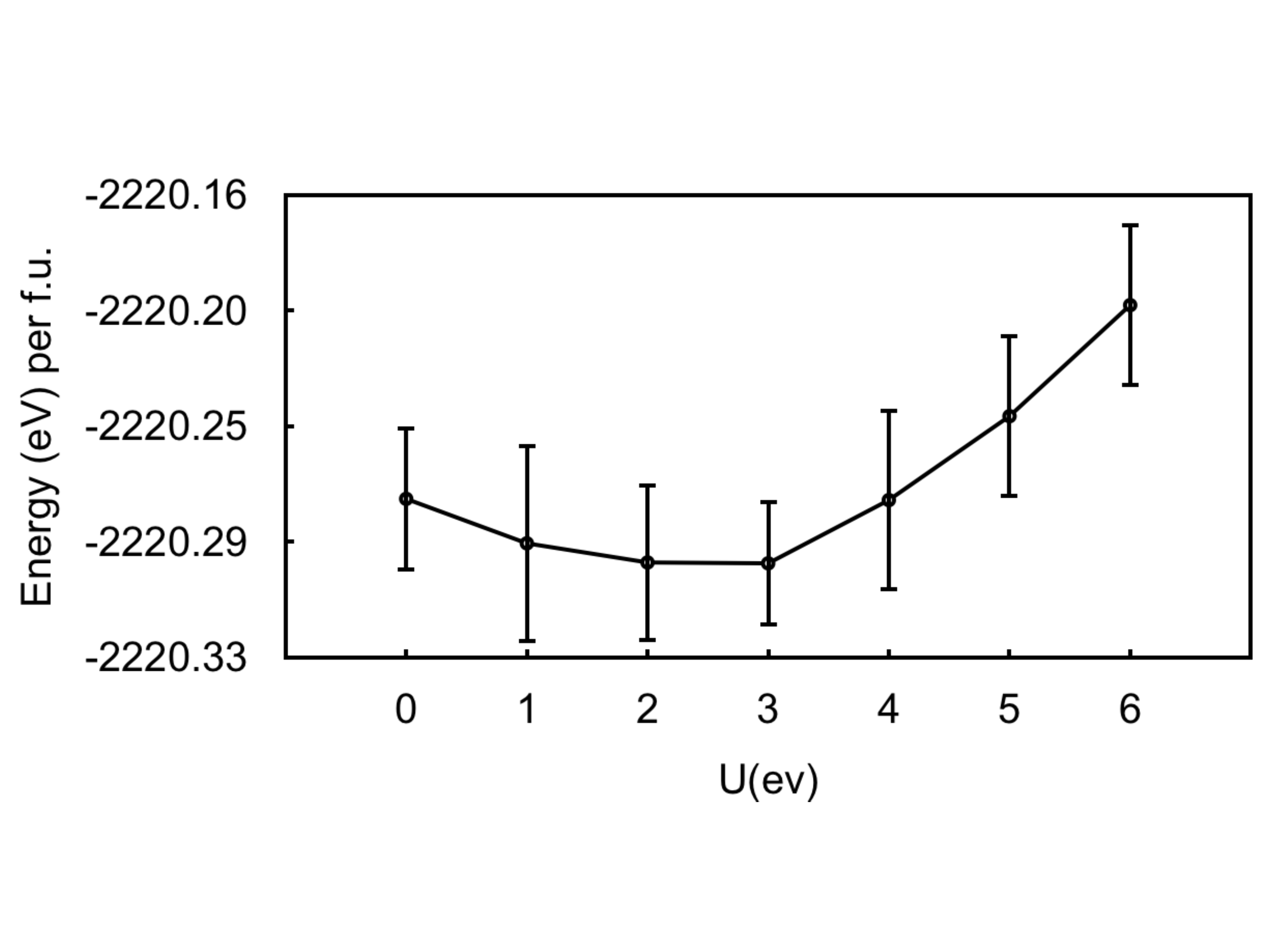}
\caption{DMC total energy of hafnia as function of the value of U.} 
\label{fig:u}
\end{figure}

\begin{figure}%[htp]
\vspace{5pt}
\includegraphics[width=0.45\columnwidth]{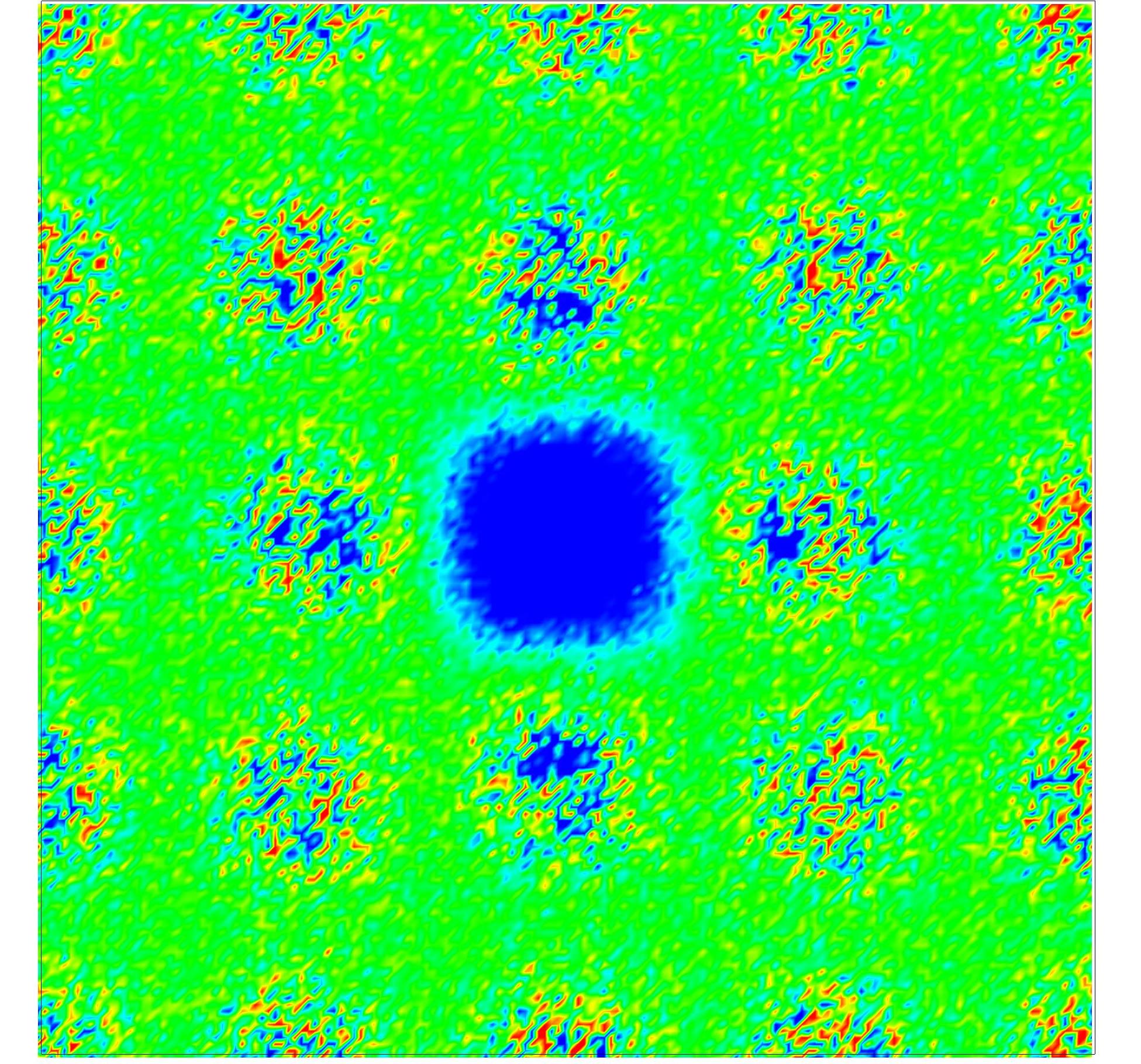}\includegraphics[width=0.45\columnwidth]{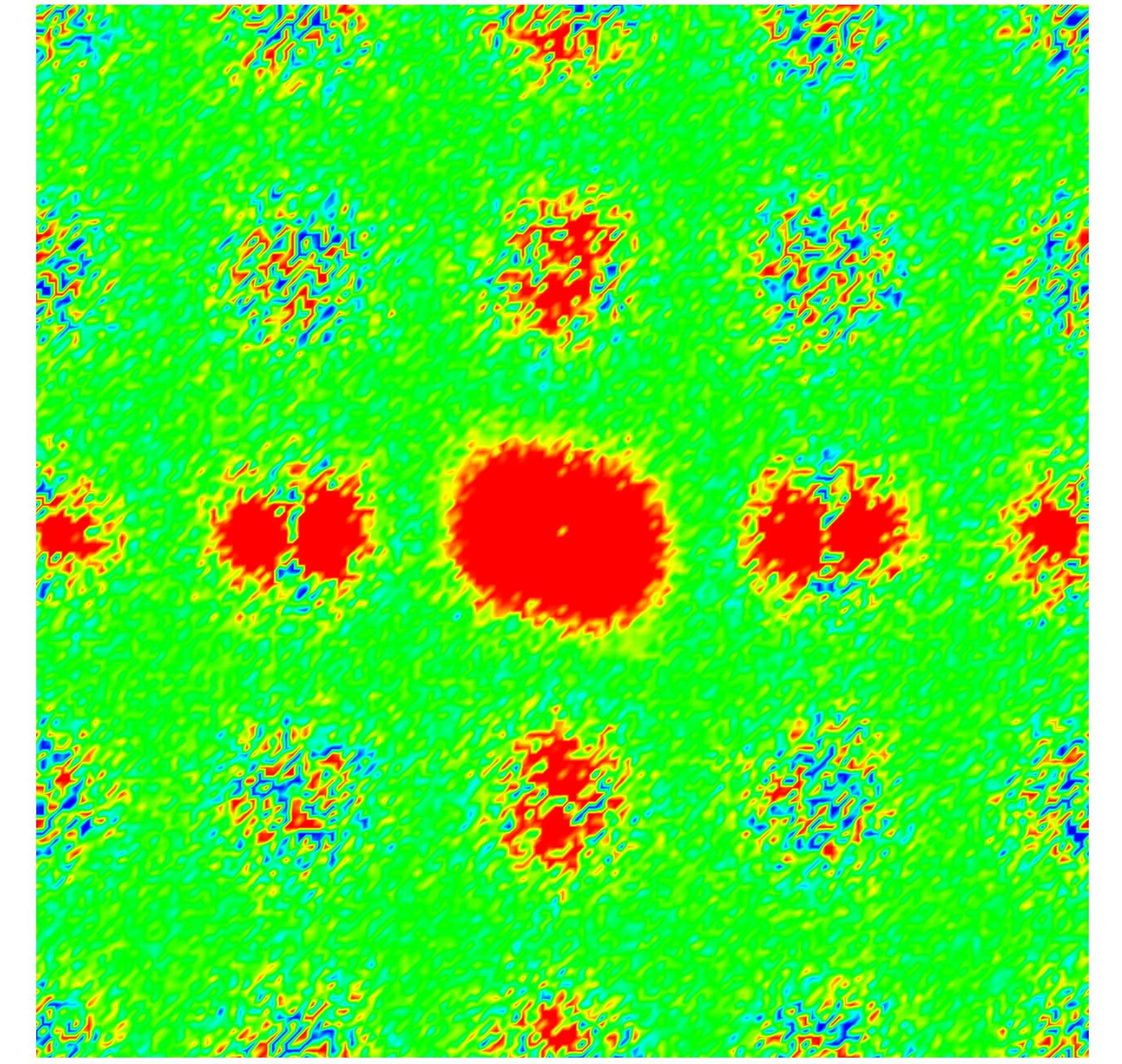}
\caption{(Color online) VMC spin densities for a charge+1 oxygen vacancy (left panel) and for an N dopant (right panel). The color scale is the same as in Fig.~\ref{fig:5}.}
\label{fig:VMC_densities}
\end{figure}
Figure~\ref{fig:5} shows that the spin densities of the charge+1 oxygen vacancy and of the N dopant have broken $C_4$ point group symmetry about the center of the defect. This is in contrast to the GGA+U results, for which the $C_4$ symmetry is preserved. An interesting question is whether the symmetry is broken at the VMC or DMC level. The trial wave functions for the VMC and DMC densities are the GGA+U wavefunction that preserve the symmetry, which suggests that correlation effects beyond GGA+U are responsible for breaking the symmetry. In Fig.~\ref{fig:VMC_densities} we show the VMC spin densities for the charge+1 oxygen vacancy (left panel) and the N-dopant (right panel) on the same color scale as Fig.~\ref{fig:5}. These panels show the ``raw" spin densities that have not been symmetrized. The figures clearly show that for the N dopant, the symmetry is broken already at the VMC level. This suggest that dynamic correlations, included in the Jastrow factor, are responsible. In contrast, the left panel suggests that the $C_4$ symmetry is preserved in VMC for the charge+1 oxygen vacancy. This implies that static correlations, that are more correctly included in DMC beyond VMC, are the primary driver to break the symmetry of the spin density for the charge+1 oxygen vacancy. Finally, we note that the maximum standard deviation in the sampled densities or spin densities at any meshpoint was no larger than about $7\times10^{-7}$~{\AA}, so that the ratio of standard deviation to modulus of density was no larger than 0.002, which is negligible.

 %\clearpage
\bibliographystyle{apsrev4-1}
\bibliography{refer}
\end{document}